\documentclass[aps,pre,reprint,superscriptaddress,amsmath,amssymb]{revtex4-1}

\usepackage{graphicx}
\usepackage[caption=false]{subfig}
\usepackage{hyperref}

\begin{document}

\title{Information geometric analysis of phase transitions in complex patterns:
the case of the Gray-Scott reaction-diffusion model}

\date{\today}

% Authors (Add full first names)
\author{Omri \surname{Har Shemesh}}
\email[Electronic mail: ]{O.HarShemesh@uva.nl}
\affiliation{Computational Science Lab, University of Amsterdam, Science Park 904,
1098XH, Amsterdam, The Netherlands}

\author{Rick Quax}
\email[Electronic mail: ]{R.Quax@uva.nl}
\affiliation{Computational Science Lab, University of Amsterdam, Science Park 904,
1098XH, Amsterdam, The Netherlands}

\author{Alfons G. Hoekstra}
\email[Electronic mail: ]{A.G.Hoekstra@uva.nl}
\affiliation{Computational Science Lab, University of Amsterdam, Science Park 904,
1098XH, Amsterdam, The Netherlands}
\affiliation{ITMO University, Saint Petersburg, Russia}

\author{Peter M.A. Sloot}
\email[Electronic mail: ]{P.M.A.Sloot@uva.nl}
\affiliation{Computational Science Lab, University of Amsterdam, Science Park 904,
1098XH, Amsterdam, The Netherlands}
\affiliation{ITMO University, Saint Petersburg, Russia}
\affiliation{Complexity Institute, Nanyang Technological University, %
60 Nanyang View, Singapore 639673, Republic of Singapore}

\begin{abstract}

  The Fisher-Rao metric from Information Geometry is related to phase
  transition phenomena in classical statistical mechanics. Several studies
  propose to extend the use of Information Geometry to study more general phase
  transitions in complex systems. However, it is unclear whether the Fisher-Rao
  metric does indeed detect these more general transitions, especially in the
  absence of a statistical model. In this paper we study the transitions
  between patterns in the Gray-Scott reaction-diffusion model using Fisher
  information.  We describe the system by a probability density function that
  represents the size distribution of blobs in the patterns and compute its
  Fisher information with respect to changing the two rate parameters of the
  underlying model. We estimate the distribution non-parametrically so that we
  do not assume any statistical model. The resulting Fisher map can be
  interpreted as a phase-map of the different patterns. Lines with high Fisher
  information can be considered as boundaries between regions of parameter
  space where patterns with similar characteristics appear. These lines of high
  Fisher information can be interpreted as phase transitions between complex
  patterns.

\end{abstract}

% PACS: 
% 64.60.an - phase transitions in finite systems
% 64.60.Bd - General theory of phase transitions
% 05.70.Fh - Phase transitions in statistical mechanics and thermodynamics
% 89.70.-a - Information theory
\pacs{64.60.Bd,64.60.an,05.70.Fh,89.70.-a}
\keywords{Information Geometry; Gray-Scott; Criticality; Fisher Information;
Complex Patterns}

\maketitle

\section{Introduction}

Phase transitions are ubiquitous in nature. They are a dramatic change in a
system's properties triggered by a minuscule shift in its environment. Phase
transitions are often associated with spontaneous symmetry breaking, where the
transition is between an unordered phase and an ordered, less symmetric
phase~\cite{Binder1987}. In simple models of phase transitions an order
parameter is defined, which is zero in the unordered phase and non-zero in the
ordered phase~\cite{Binder1987,ZinnJustin1989,Hohenberg2014}. This is the basis
for a mean-field approach to phase transitions, where an expansion of the free
energy in the order parameter is performed. Many powerful methods have been
developed over the years to study phase transitions, especially in the study of
universality in so-called second order (or critical) transitions, such as the
Landau-Ginzburg theory~\cite{Hohenberg2014,Yukhnovskii1987} and Wilson's
Renormalization Group approach~\cite{Wilson1974,Binder1987,ZinnJustin1989}. For
those transitions the order parameter is continuous, thermodynamic quantities
obey scaling laws in the vicinity of the critical point, and there is a
diverging correlation length. For first order transitions, on the other hand,
there is a jump in the value of the order parameter, there is no diverging
correlation length and thus no scaling of the thermodynamic functions near the
transition point. During the transition there can be a mixed phase with a
stable interface between the two phases~\cite{Binder1987}. The study of first
order transitions is very important for complex systems, especially social or
ecological complex systems, because the sudden jump between two phases (which
is discontinuous) can be quite dramatic~\cite{VillaMartin2015}.

While the Ginzburg-Landau-Wilson approach has been tremendously successful in
explaining universality in second-order phase transitions, it requires the
definition of an order parameter. Different approaches, which do not require an
order parameter, can be useful for cases where an order parameter is difficult
to identify, or does not exist (e.g.~\cite{Wegner1971}). In one such approach
we study the probabilistic description of the system while changing the
parameters to bring the system across a transition. The statistical properties
of the system in the different phases are very different. Therefore, at the
phase transition the shape of the probability distribution function will change
drastically. This is captured by the Fisher information matrix (FIM) through
the Cram\'{e}r-Rao bound~\cite{Fisher1922,Cramer1946,Rao1945,Cover2006}. A
compelling differential-geometric framework to study the changes that the
probability distribution undergoes is Information Geometry (IG)~\cite{Amari}.
In IG the family of probability distributions that are parametrized by a set of
continuous parameters (designated here as the vector $\theta$) is seen as a
differential manifold. The parameters form a coordinate system on the manifold,
and distances are measured by the Fisher-Rao metric:
\begin{equation}
    \label{eq:fisher_rao}
    g_{\mu\nu}(\theta) = \langle \partial_\mu \ln p\, \partial_\nu \ln p \rangle
\end{equation}
which is a positive semi-definite, symmetric matrix that changes covariantly
under reparametrizations of the probability distribution $p(x;\theta)$. Here
$\partial_\mu \equiv \partial/\partial\theta^\mu$ is a derivative with respect
to one of the parameters, indexed by $\mu$.  The IG of many models in
statistical mechanics has been studied, e.g., in~\cite{Ingarden,%
Ingarden1982,Janyszek1989,Janyszek1999,Janyszek1990a,Ruppeiner1979,%
Ruppeiner1981,Ruppeiner1995,Ruppeiner1990,Ruppeiner2012,Brody1995,Brody2003,%
Brody2009,Janke2004}. Of particular interest in these studies is the role of
the scalar (Riemannian) curvature. It was shown to diverge at critical
transition points and on the spinodal curve~\cite{Brody2009,Janke2004}, thus
effectively preventing geodesics from crossing into the unphysical area of
phase space~\cite{Kumar2012}. See also~\cite{Maity2015} for a general
renormalization group analysis of IG near criticality.

A connection between phase transitions in IG and the Ginzburg-Landau-Wilson
approach can be made when an order parameter $\phi_\mu$ is the derivative of a
thermodynamic potential with respect to some thermodynamic variable
$\theta^\mu$. Then there exists a collective variable $X_\mu(x)$ such that
$\phi_\mu = -k_B T \langle X_\mu \rangle$~\cite{Yukhnovskii1987,Prokopenko2011}
and the Fisher information matrix can be shown to obey~\cite{Prokopenko2011}:
\begin{equation}
    \label{eq:fisher_order_params}
    g_{\mu\nu} = - \frac{\partial\langle X_\mu \rangle }{\partial\theta^\nu}%
    = \beta \frac{\partial \phi_\mu}{\partial \theta^\nu}
\end{equation}
where $\beta=(k_BT)^{-1}$ is the inverse temperature, $k_B$ being Boltzmann's
constant.  At second order phase transitions in the thermodynamic limit this
derivative diverges and therefore a corresponding entry of the FIM also
diverges.  

When the system is finite, the Fisher information does not diverge but rather
attains a maximum. The maximum of the Fisher information has been used to
accurately find the phase transition point in finite
systems~\cite{Wang2011,Prokopenko2011} and as a definition of criticality in
living systems~\cite{Hidalgo2014}.

We have two main goals for the current work: first to test a conjecture set
forth by Prokopenko \emph{et al.} in~\cite{Prokopenko2011} that a divergence
(or maximization) of the entries of the FIM can detect phase transitions even
in the absence of an order parameter. 

Second, to measure the Fisher information matrix without resorting to the
underlying dynamics of the system and without assuming a specific parametric
model for equation~\eqref{eq:fisher_rao}. This addresses the problem that often
the microscopic dynamics of complex systems are unknown, and an analytic
description of the probability density function is missing.

To accomplish these two goals we chose to study the specific example of the two
dimensional Gray-Scott (GS) reaction diffusion model~\cite{Gray1984}. We chose
the GS model for its rich variety of spatial and spatio-temporal
patterns~\cite{Pearson1993} and we consider the transitions between the
different patterns as critical transitions.  Among the different types of
patterns one can find self-replicating
spots~\cite{Reynolds1994,Wei2001,Wei2003}, spatio-temporal
chaos~\cite{Wang2007}, and labyrinthine patterns~\cite{Jones1998}.  These were
first systematically classified by Pearson~\cite{Pearson1993}. Our goal is to
use the Fisher information matrix to construct a phase map for the Gray-Scott
model, where we expect areas with high values of the Fisher information to
demarcate the different patterns. As a probabilistic description for the system
we chose the blob-size distribution, which we take to be a function of the
control parameters of the model $F$ and $k$ and which we estimate
non-parametrically by using image processing on the resulting spatial $V$
concentration from our simulations.

The paper is organized as follows: in Section~\ref{sec:fisher_criticality} we
discuss the relationship between Fisher information and criticality, which
forms the motivation for our approach. We revisit the arguments
in~\cite{Prokopenko2011} and extend them to our case. In Section~\ref{sec:gs}
we introduce the Gray-Scott model and discuss some of its properties. In
Section~\ref{sec:results} we present the results of the computations and in
Section~\ref{sec:methods} we explain the methods we used to compute the
Fisher information in this settings. Last we discuss our results in
Section~\ref{sec:conclusions}.

%%%%%%%%%%%%%%%%%%% FISHER INFORMATION AND CRITICALITY %%%%%%%%%%%%%%%%%%%%%%%
\section{Fisher Information and Criticality}
\label{sec:fisher_criticality}

In this section we summarize the derivation performed in~\cite{Prokopenko2011}
leading to Eq.~\eqref{eq:fisher_order_params} that relates order parameters and
the entries of the Fisher information matrix. This derivation leads to the
conjecture~\cite{Prokopenko2011} that it is enough to consider the Fisher
information matrix entries rather than order parameters and is presented here
because it is important for our exposition.

The Gibbs ensemble can be generically written in the following way:
\begin{equation}
    \label{eq:gibbs_ensemble}
    p(x;\theta) = \frac{1}{Z(\theta)}\exp \left[ -\theta^\mu
    X_\mu(x) \right]
\end{equation}
with $Z(\theta)$ set by normalization and with summation convention over
repeated indices, which we use throughout the paper. For this distribution, the
Fisher information Eq.~\eqref{eq:fisher_rao} is~\cite{Prokopenko2011}:
\begin{equation}
    \label{eq:fisher_gibbs}
    g_{\mu \nu}(\theta) = \partial_\mu\partial_\nu\ln Z(\theta) =
    \partial_\mu\partial_\nu (-\beta G)
\end{equation}
where $G$ is the Gibbs free energy. Performing the derivatives we obtain
Eq.~\eqref{eq:fisher_order_params}.

For many systems, the order parameter can be defined by introducing an external
field $h$ to the free energy that couples to the order parameter
$\phi$~\cite{Binder1987}. The canonical example being the magnetization that
couples to the external magnetic field, so that $M=-\left(\frac{\partial
A}{\partial h}\right)_T$. The external field being one of the external
parameters $\theta^\mu$. We then have:
\begin{equation}
    \label{eq:order_parameter}
    \phi = -{\left( \frac{\partial G}{\partial h} \right)}_T.
\end{equation}
Setting $\theta^{\hat{\mu}}=h$ for a particular $\hat{\mu}$ we obtain:
\begin{equation}
  \label{eq:order_parameter_theta}
  \phi_{\hat{\mu}} = -\frac{\partial G}{\partial \theta^{\hat{\mu}}} = -k_B T \langle X_{\hat{\mu}} \rangle.
\end{equation}
The meaning of Eq.~\eqref{eq:order_parameter_theta} is that if an order
parameter $\phi_{\hat{\mu}}$ is a derivative of the free energy $G$, then there
exists a collective variable $X_{\hat{\mu}}(x)$ whose average is proportional
to the order parameter~\cite{Prokopenko2011}. It is important to note that many
models exist whose order parameter is indeed the average of a collective
variable~\cite{Prokopenko2011,Yukhnovskii1987}.

Combining equations~\eqref{eq:fisher_gibbs}
and~\eqref{eq:order_parameter_theta} we obtain Eq.~\eqref{eq:fisher_order_params}.
This links the value of the entries of the Fisher information with the
derivatives of the order parameters $\phi_\mu$ of the system. Since at phase
transitions the order parameter or its derivatives becomes non-analytic we can
expect the Fisher information matrix to have diverging entries at the phase
transition point. For example, at a ferromagnetic transition point, with
$(\theta^1, \theta^2) = (h, T)$, the diagonal elements of the Fisher information
matrix are given by~\cite{Prokopenko2011}:
\begin{align}
    \label{eq:fisher_chi}
    k_B T g_{11}(h) &= {\left( \frac{\partial \phi}{\partial h} \right)}_T =
    -{\left(\frac{\partial^2G}{\partial h^2}\right)}_T \equiv \chi_T\,; \\
    \label{eq:fisher_ch}
    k_B T g_{22}(T) &= {\left( \frac{\partial S}{\partial T} \right)}_h =
    -{\left(  \frac{\partial^2 G}{\partial T^2} \right)}_h \equiv
    \frac{C_h}{T}\,.
\end{align}
Since $g_{11}$ and $g_{22}$ are proportional to the magnetic susceptibility
$\chi_T$ and heat capacity $C_h$ respectively, we expect both entries to diverge
at the point of the magnetic second order phase
transition~\cite{Prokopenko2011}. More generally, it is easy to show
that~\cite{Prokopenko2011}
\begin{equation}
    \label{eq:fisher_variance}
    g_{\mu \nu}(\theta) = \langle X_\mu X_\nu \rangle - \langle X_\mu \rangle
    \langle X_\nu \rangle.
\end{equation}
This is also called the generalized susceptibility in statistical
mechanics~\cite{Hidalgo2014}. This derivation led Prokopenko to propose that
the maximization of the appropriate Fisher information matrix can detect phase
transitions, without explicitly defining an order
parameter~\cite{Prokopenko2011}.

The relation~\eqref{eq:fisher_order_params} suggests the introduction of an
order parameter derived from the Fisher information by integrating it from one
phase to the next, in the following way:
\begin{equation}
    \label{eq:fisher_order_param}
    \int\limits_{\theta_1}^{\theta_2} g_{\mu\nu}(\theta)d\theta^\nu =
    \int\limits_{\theta_1}^{\theta_2}\frac{\partial\phi_\mu}{\partial\theta^\nu}
    d\theta^\nu = \int\limits_{\theta_1}^{\theta_2} d\phi_\mu =
    \phi_\mu(\theta_1,\theta_2).
\end{equation}
We absorbed the inverse temperature in the definition of $\theta$, and the
integration path starts at $\theta_1$ which is in one phase and ends at
$\theta_2$ which is in the other phase. In general this will depend on the
integration path and the end-points.

While the derivation above assumes the form~\eqref{eq:gibbs_ensemble} for the
probabilistic description of the system, the idea of Fisher information
maximization at phase transition points can be generalized for other
probabilistic descriptions based on the Cram\'{e}r-Rao
bound~\cite{Cramer1946,Rao1945,Cover2006} that states that the variance an
unbiased estimator $\hat{\theta}(x)$ is bounded from below by the inverse of
the Fisher information. We make the following heuristic argument: when a system
is said to undergo a phase transition, it means that there is an observable
change in some aspect of the system (often having to do with the symmetries of
the system). This means that the statistical properties of the system in the
two phases differ significantly.  For example, if we sample the energy per spin
of an Ising spin system repeatedly in the high-temperature phase, we will
obtain a broad distribution of energies. Conversely, at the low-temperature
phase the energy distribution is very narrow, since in the low-temperature
phase the spins are aligned and the system is in the ground
state~\cite{Wilson1974}.  Thus, if we look at the probability density function
describing these observables change as a function of the control parameter, it
undergoes a drastic change in its functional form. This, in turn, implies that
we can estimate the value of the control parameter at the phase transition
point accurately, because of the large change in the behavior of the density
function. According to the Cram\'er-Rao inequality, the inverse of the value of
the Fisher information serves as a lower bound on the variance of the estimated
parameter. If this parameter can be estimated accurately then this implies a
high value of the Fisher information. We therefore surmise that under very
general circumstances the Fisher information is maximized at phase transition
points. 

%%% THE GRAY-SCOTT MODEL
\section{The Gray-Scott Model}
\label{sec:gs}

The Gray-Scott model is a non-linear reaction-diffusion model of two chemical
species $U$ and $V$ with the reactions
\begin{align}
    \label{eq:gs_model}
U + 2V &\rightarrow 3V \\
V&\rightarrow P.\nonumber
\end{align}
$U$ is constantly supplied into the system and the inert product $P$ removed.

We can simulate the reaction using the law of mass action~\cite{Hinrichsen2000},
where we assume that the rate of each reaction is proportional to the
concentration of the reactants at each point. The resulting non-linear coupled
differential equations are:
\begin{align}
  \label{eq:gs_diff}
  \frac{\partial u}{\partial t} &= D_u \nabla^2 u - uv^2 + F(1-u) \\
  \frac{\partial v}{\partial t} &= D_v \nabla^2 v + uv^2 - (F+k)v \nonumber
\end{align}
where $u = u(t, \mathbf{x})$, $v = v(t, \mathbf{x})$ are the (dimensionless)
concentrations of the two chemical species, $\nabla^2$ is the Laplacian with
respect to $\mathbf{x}$, $D_u$ and $D_v$ are diffusion coefficients of $u$ and
$v$ respectively, $F$ represents the rate of the feed of $U$ and the removal of
$U$, $V$ and $P$ from the system and $k$ is the rate of conversion of $V$ to
$P$. In practice this is a model of chemical species in a gel reactor where the
rate $F$ can be relatively easily modified, and $k$ is dependent on the
temperature of the system. $D_u$ and $D_v$ are more difficult to change and we
will consider them constant, with $D_u/D_v = 2$.

\subsection{Linear stability analysis}
\label{sub:linear_stability_analysis}
\begin{figure}[h!]
  \centering
  \includegraphics[width=\columnwidth]{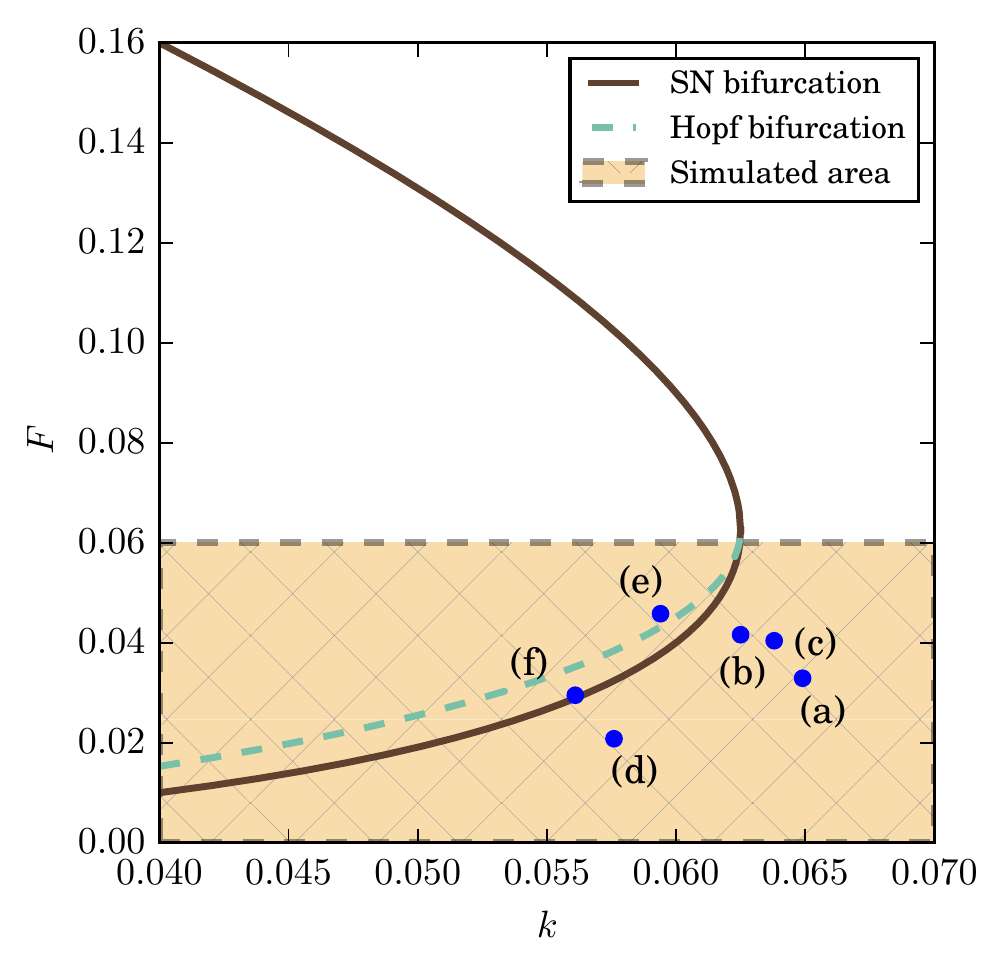}
  \caption{Bifurcation diagram of the Gray-Scott model. The solid line
  indicates the saddle-node bifurcation and the dashed line the Hopf
  bifurcation. The highlighted area is where we perform our simulations.
  The blue dots are the position in parameter space of the patterns that appear 
  in Fig.~\ref{fig:patterns}.}\label{fig:bifurcation}
\end{figure}

We start with the standard stability analysis for the homogeneous system (i.e.\
without diffusion). The Gray-Scott model has a trivial homogeneous steady state
solution, referred to as the red state, at $[u_R, v_R] = [1,0]$ which is always
linearly stable for positive $F$ and $k$~\cite{Pearson1993,Mazin1996}. Under the
condition that
\begin{equation}
    \label{eq:discriminant}
    d = 1 - 4(F+k)^2/F > 0
\end{equation}
two additional homogeneous steady states solutions appear:
\begin{equation}
    \label{eq:blue_state}
    [u_B, v_B] = \left[\frac{1}{2}(1-\sqrt{d}),
    \frac{1}{2}\frac{F}{F+k}(1+\sqrt{d})\right]
\end{equation}
\begin{equation}
    \label{eq:i_state}
    [u_I, v_I] = \left[\frac{1}{2}(1+\sqrt{d}),
    \frac{1}{2}\frac{F}{F+k}(1-\sqrt{d})\right].
\end{equation}
These are referred to as the blue state and the intermediate state
respectively~\cite{Mazin1996}. Linear stability analysis  shows that when these
states exist, the intermediate state is always unstable whereas the blue state
can be stable. The two states appear through a saddle-node bifurcation which
in the $F-k$ plane is defined by the curve:
\begin{equation}
    \label{eq:SN_bifurcation}
    F_{SN}(k) = \frac{1}{8}\left[ 1 - 8k \pm \sqrt{1-16k} \right].
\end{equation}
Under certain conditions~\cite{Mazin1996} the blue state undergoes a Hopf
bifurcation at
\begin{equation}
    \label{eq:Hopf_bifurcation}
    F_{H}(k) = \frac{1}{2}\left[ \sqrt{k} -2k -\sqrt{k(1-4\sqrt{k})} \right].
\end{equation}
For more details see~\cite{Mazin1996,Wei2001,Wei2003,McGough2004}.
The bifurcation curves are plotted in Fig.~\ref{fig:bifurcation} together with
the area where we performed our simulations.

\subsection{Complex patterns}
\label{sub:complex_patterns}

\begin{figure*}[ht!]
  \subfloat[Self-replicating spots after replication is finished. $F=0.0392,%
    k=0.0649$ (Pearson $\lambda$ pattern).\label{fig:bacteria}]{%
    \includegraphics[width=0.3\textwidth]{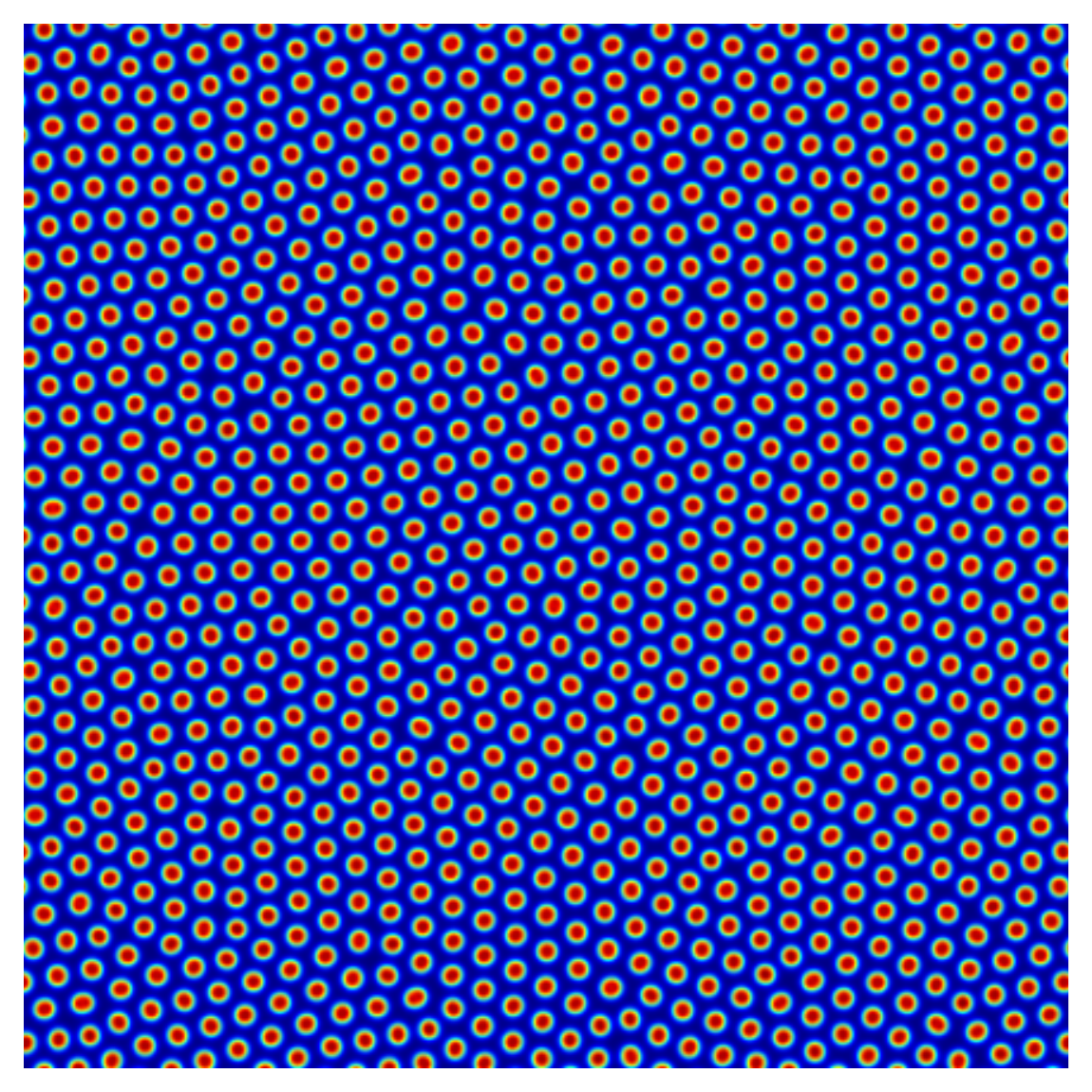}}
  \subfloat[Worm-like patterns. $F=0.0416, k=0.0625$
  (Pearson $\mu$ pattern).\label{fig:worms}]{%
    \includegraphics[width=0.3\textwidth]{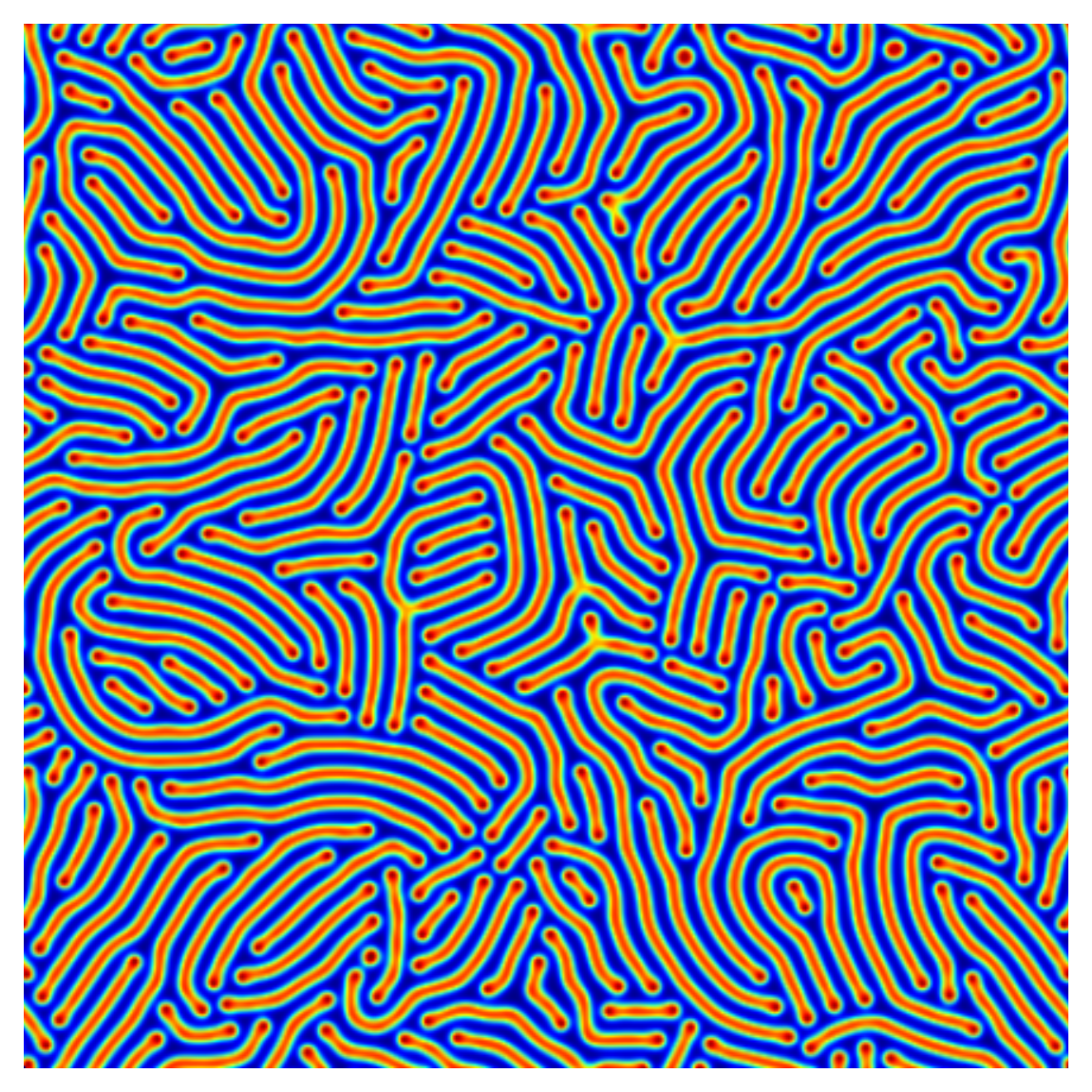}
   }
   \subfloat[``Mixed-phase'' between the self-replicating spots and the
   worm-like pattern. $F=0.0404, k=0.0638$
   (Pearson $\eta$ pattern).\label{fig:mixed_phase}]{%
    \includegraphics[width=0.3\textwidth]{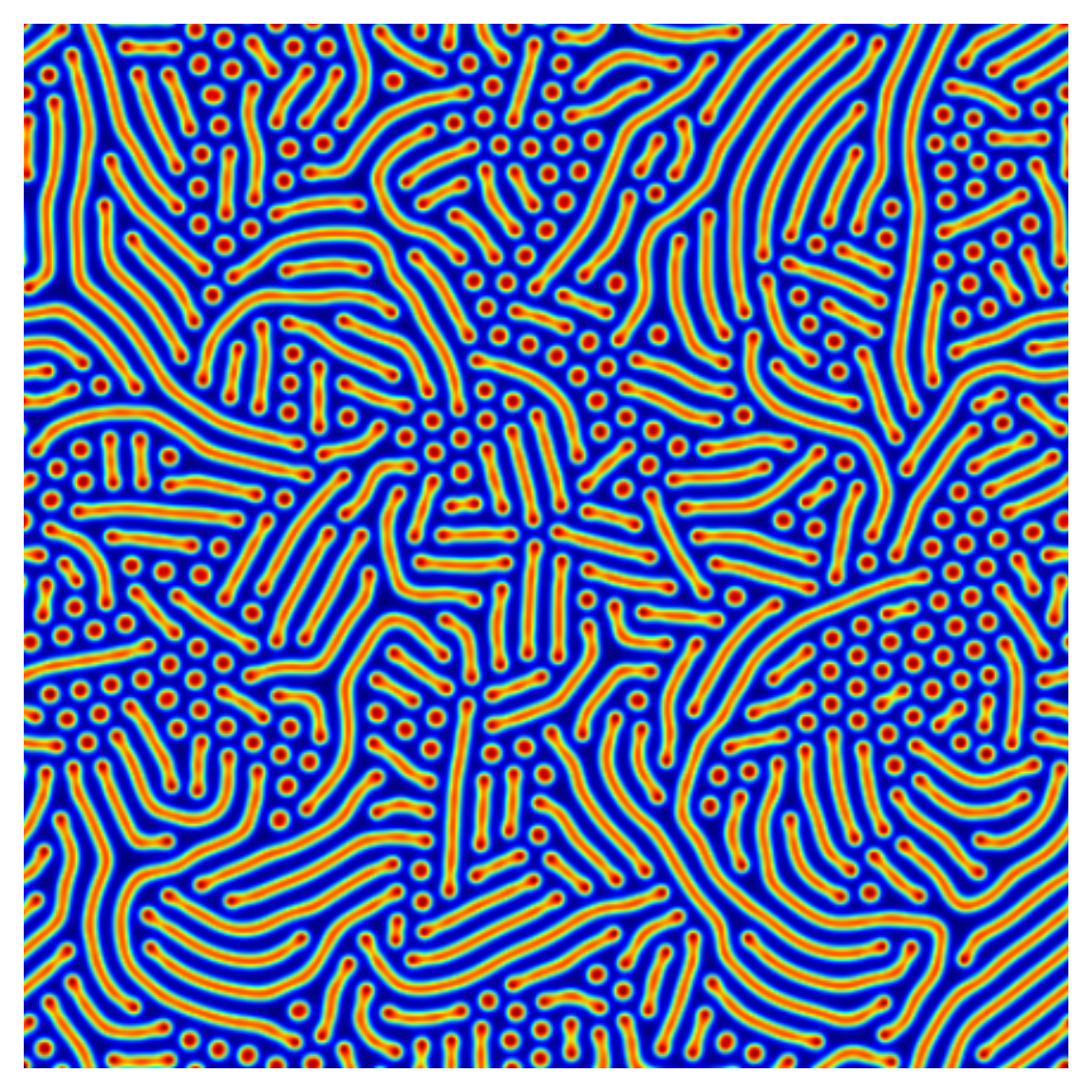}
    }\\
    \subfloat[Spatio-temporal chaos (never reaches steady-state).$F=0.0208, k=0.0576$
    (Pearson $\epsilon$ pattern).\label{fig:chaos}]{%
    \includegraphics[width=0.3\textwidth]{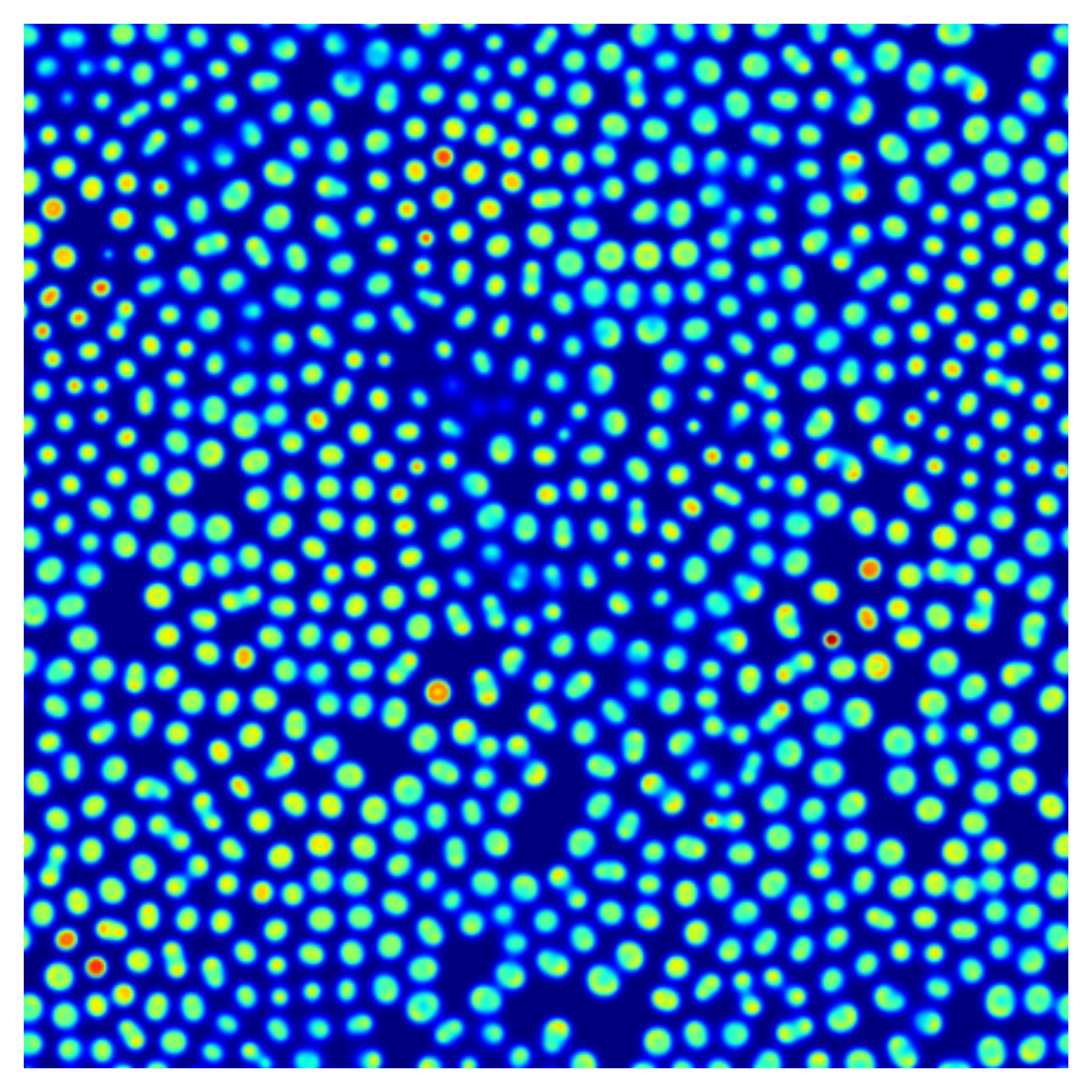}}
    \subfloat[$F=0.0175, k=0.0504$ (Pearson $\alpha$ pattern).\label{fig:crazy}]{%
      \includegraphics[width=0.3\textwidth]{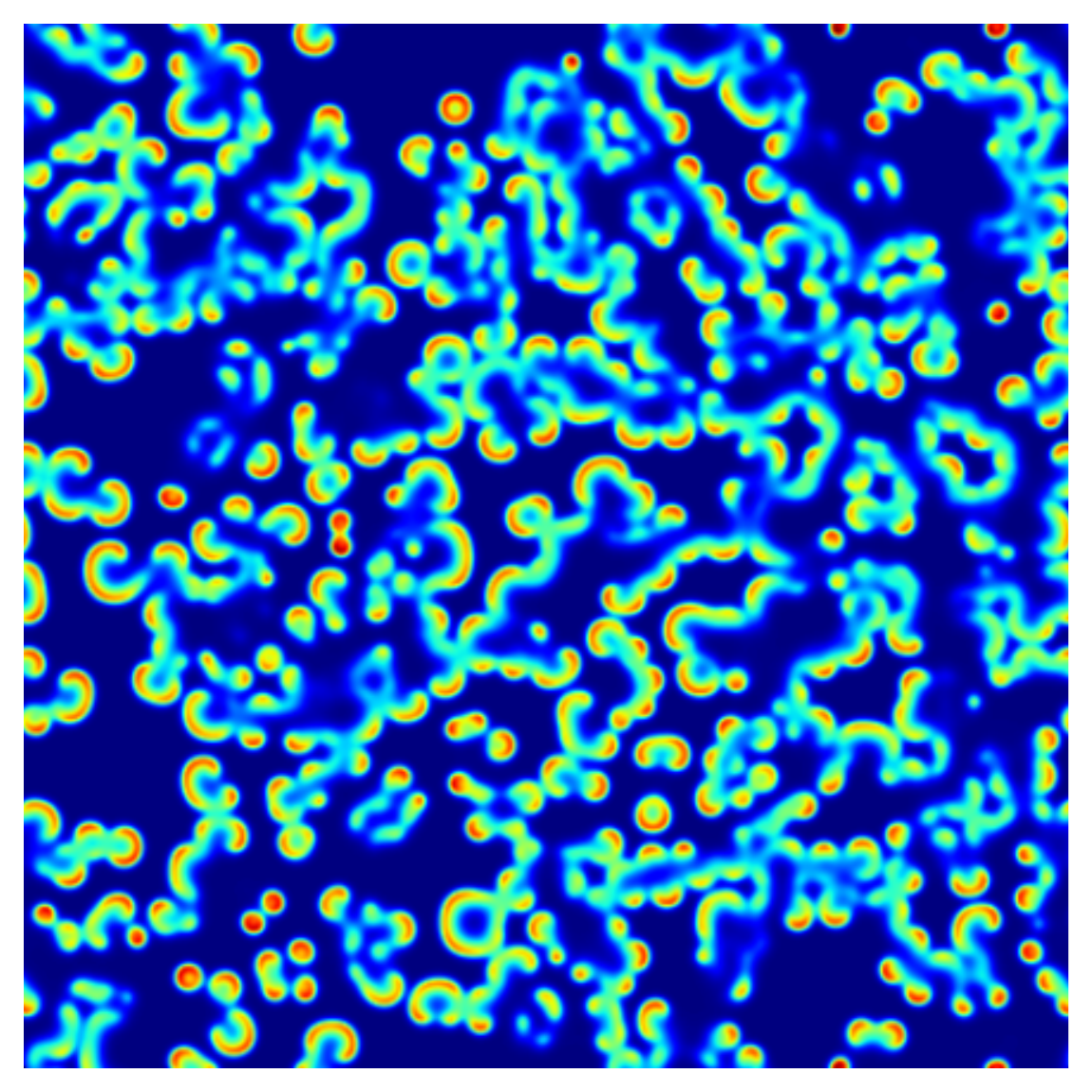}}
    \subfloat[Labyrinthine pattern. $F=0.0295, k=0.0561$
    (Pearson $\theta$ pattern).\label{fig:labyrinthine}]{%
      \includegraphics[width=0.3\textwidth]{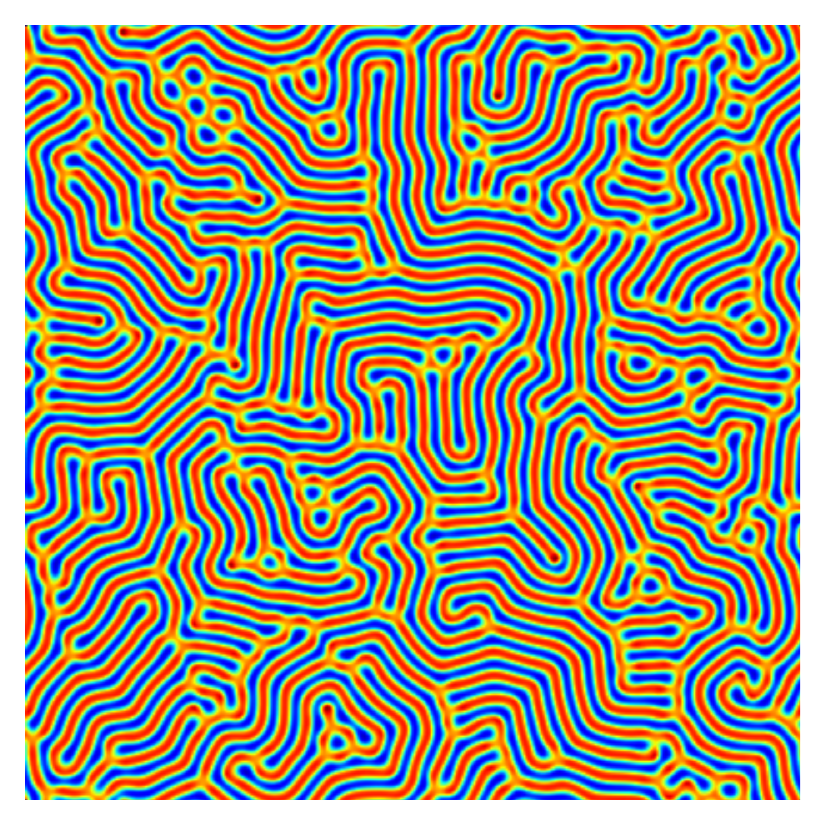}}
  \caption{Example of patterns found in the Gray-Scott model. These represent the
  concentration of $V$. All simulations performed with a grid size of
  $400\times 400$, and $D_u/D_v=2$.}\label{fig:patterns}
\end{figure*}

In the vicinity of the bifurcations a variety of inhomogeneous patterns may
appear~\cite{Mazin1996}. These can be observed by setting the initial state of
the system to the red state and adding a finite perturbation that allows the
system to reach a different attractor (for details see Sec.~\ref{sec:methods}).
The first to systematically study these patterns in two dimensions was
Pearson~\cite{Pearson1993}. Pearson classified the patterns in $12$ types and
designated them with Greek letters.  In this paper we use the same parameters
Pearson used for his simulations, except that we will vary the simulation
lattice size to include larger patterns, as will be described later. The
different patterns appear close to each other in the $F-k$ space and often mix
on the boundaries of regions of different patterns. One of the patterns that
was first discovered by Pearson~\cite{Pearson1993} is that of the
self-replicating spots (e.g.\ Fig.~\ref{fig:bacteria}). A single spot of high
$V$ with a well defined boundary grows until at a certain point it will split
into two spots. The process continues until the whole simulation area is
covered with spots~\cite{Reynolds1994}. Depending on the parameters, the
self-replicating spots will either reach an asymptotic fixed point or will
start to flicker in what is known as spatio-temporal
chaos~\cite{Reynolds1994,Nishiura1999,Nishiura2001,Wei2001,Wei2003}.  Wang and
Ouyang~\cite{Wang2007} derived a probabilistic description of the spot count in
the chaotic regime as a function of the rate of spot creation and annihilation
which were found to linearly depend on the spot-count. Another group of
patterns are the worm-like patterns (Fig.~\ref{fig:worms}) which grow and fill
the entire simulation lattice and then are fixed. In the parameter space,
between the stable self-replicating spots and the worm-like patterns, a mixed
pattern appears which contains both spots and stripes. This pattern
(Fig.~\ref{fig:mixed_phase}) is reminiscent of a first order phase transition,
where phase co-existence appears in the transition region. The degree of mixing
gradually increases and then decreases again across the transition.

In the examples in Fig.~\ref{fig:patterns}, (a), (b), (c) and (f) patterns are
at a fixed point of the dynamics. The patterns (d) and (e), on the other hand,
are a snapshot of patterns that do not reach a fixed-point. The patterns that
appear in the Gray-Scott are examples of self-organization because they are
macroscopic patterns that appear through the local interactions of the $U$ and
$V$ substances~\cite{Prokopenko2009a}.

\subsection{Probabilistic description}
\label{sub:probabilistic_description}

\begin{figure*}[ht!]
  \subfloat[Original image.\label{fig:orig_process}]{%
    \includegraphics[width=0.3\textwidth]{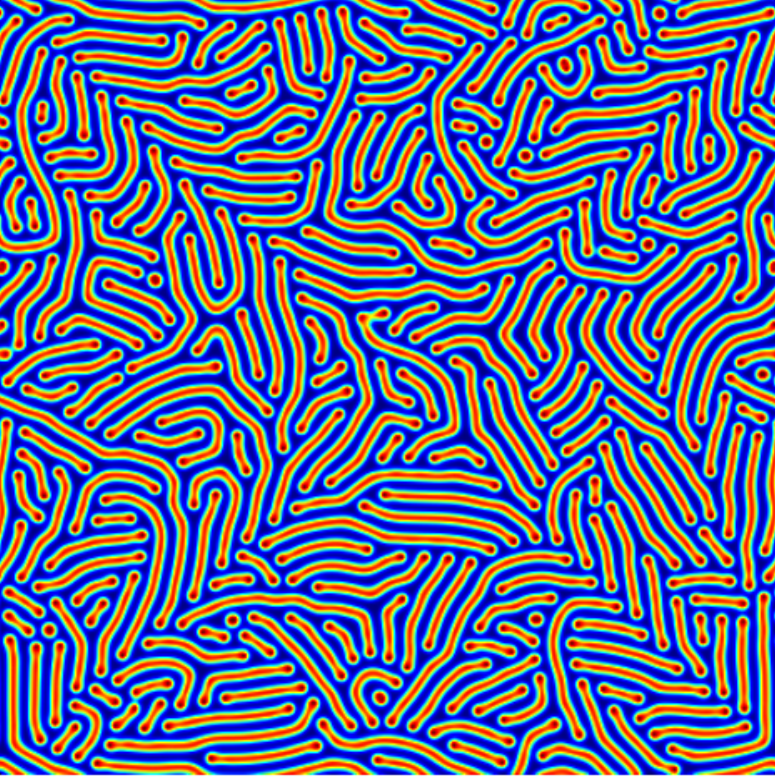}}
  \subfloat[Image with detected blobs superimposed.\label{fig:with_blobs}]{%
    \includegraphics[width=0.3\textwidth]{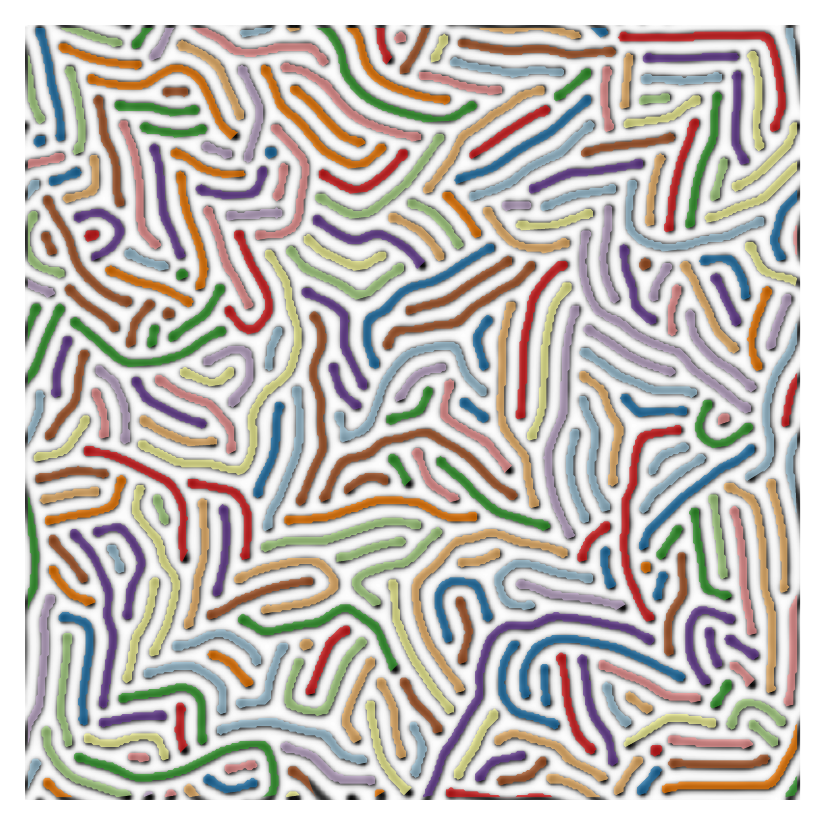}}
  \subfloat[Histogram of blob areas with density estimation.\label{fig:histogram}]{%
    \includegraphics[width=0.3\textwidth]{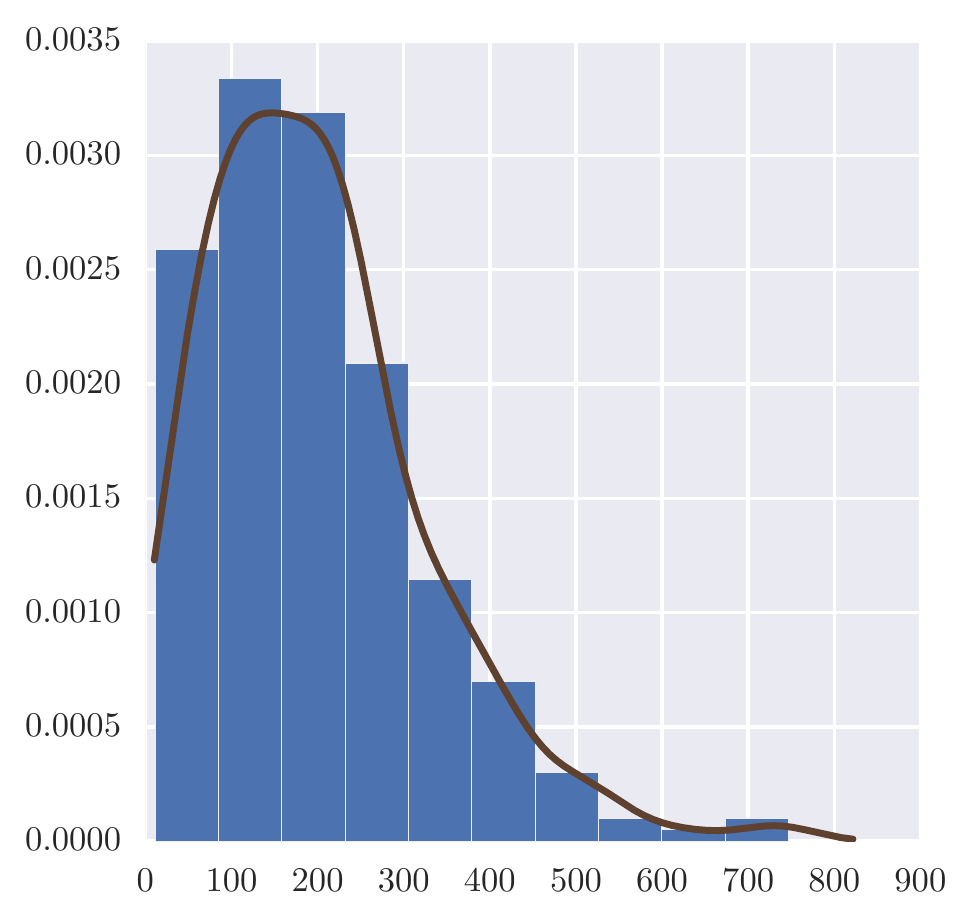}
  }\\
  \subfloat[Original image.\label{fig:orig_process2}]{%
  \includegraphics[width=0.3\textwidth]{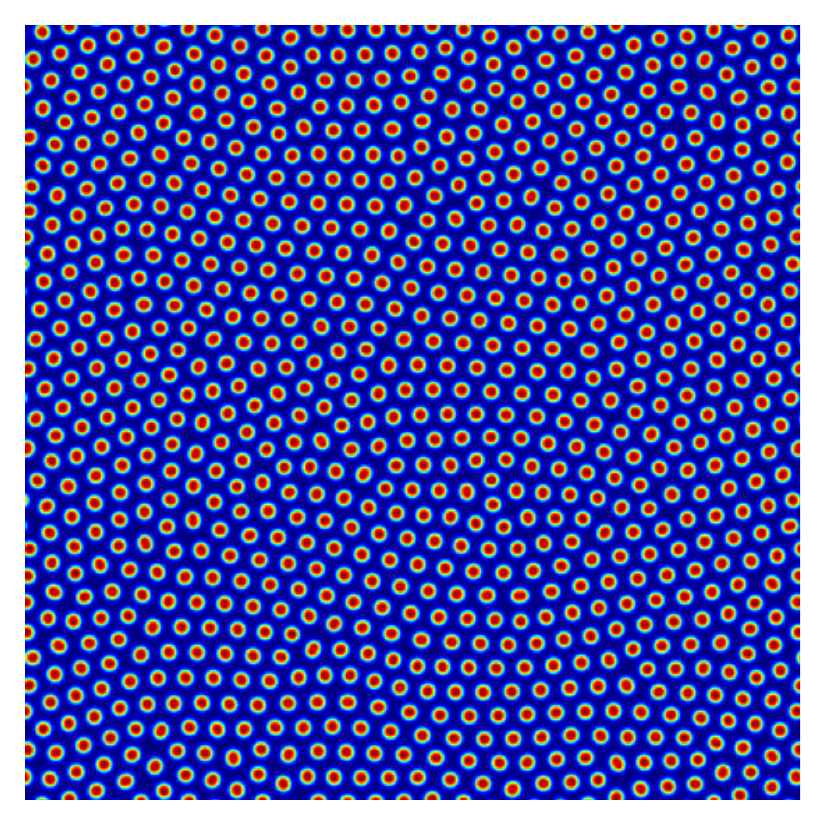}}
  \subfloat[Image with detected blobs superimposed.\label{fig:with_blobs2}]{%
    \includegraphics[width=0.3\textwidth]{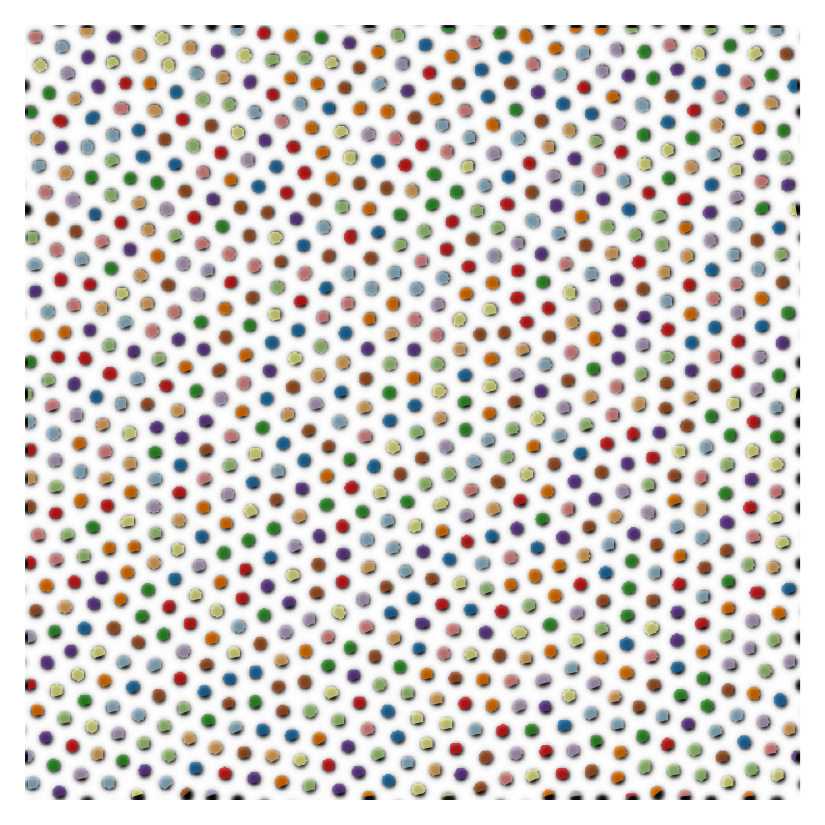}
  }
  \subfloat[Histogram of blob areas with density estimation.\label{fig:histogram2}]{%
    \includegraphics[width=0.3\textwidth]{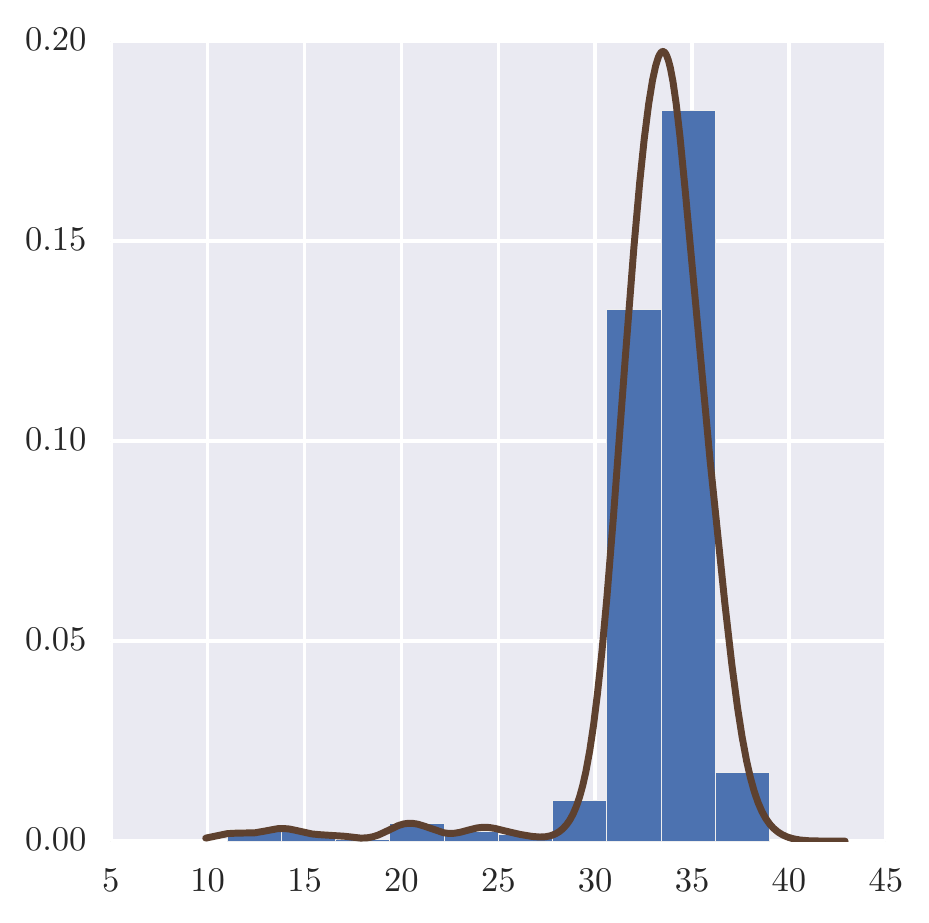}
  }
  \caption{Two examples of the density extraction process. The original images~\ref{fig:orig_process}
  and~\ref{fig:orig_process2} are binarized and a blob detection algorithm is run.
  This results in a set of blobs (depicted in~\ref{fig:with_blobs} and~\ref{fig:with_blobs2}) from
  which a histogram of blob sizes is computed and a Gaussian kernel density
  estimation is used to obtain the final density function (plotted in~\ref{fig:histogram}
  and~\ref{fig:histogram2}). First picture at $F = 0.0413;\;
  k = 0.0628$ and second picture at $F = 0.0335;\; k = 0.0644$.}\label{fig:extraction_process}
\end{figure*}

In order to construct a phase map using the Fisher information matrix, we
follow Wang~\cite{Wang2007} in constructing our probabilistic description of
the patterns. Wang treated each spot as an entity and computed the probability
for a number of spots to appear at a unit time. We separate the pattern into
similar entities, but regard not only the self-replicating spots but also the
stripes and the labyrinthine patterns as entities, which we call ``blobs''.  We
take the lattice of $V$ values (the $U$ value lattice is usually very similar
to the $V$ one) and treat it as an image. To identify a blob we first binarize
the image using Otsu's method\cite{Otsu1975}. We use the \verb;find_blobs;
method from the image processing python library
\verb;SimpleCV;~\footnote{\url{http://simplecv.org}} to label continuous
clusters in the binarized image. We extract the size of each blob (in pixels)
and use a non-parametric estimation procedure to obtain a PDF of the sizes of
the blobs.  We assume that the blob sizes are characteristic of the different
patterns and that by observing their distribution we could, with a high degree
of certainty, deduce the type of pattern. We therefore expect the Fisher
information to become large when a PDF changes rapidly and therefore we will be
able to detect the transitions between the patterns. The density extraction
process is depicted in Fig.~\ref{fig:extraction_process}.

We assume that the extracted PDF is a function of the parameters $F$ and $k$
and that we can therefore compute derivatives of it with respect to these
parameters by using a finite difference scheme. We use a centered finite
difference scheme and the integration is performed by use of
\verb;scipy.integrate.quad;~\footnote{\url{http://www.scipy.org}}. The
expression we use for the Fisher information is:
\begin{align}
\label{eq:fisher_finite_diff}
g_{\mu\nu}(\theta) &= \int \frac{\ln p(x|\theta+\Delta\theta^\mu) - \ln p(x|\theta-\Delta\theta^\mu)}{2\Delta\theta^\mu} \times \\
&\times \frac{\ln p(x|\theta+\Delta\theta^\nu) -
\ln p(x|\theta-\Delta\theta^\nu)}{2\Delta\theta^\nu} p(x;\theta)\, dx.
\nonumber
\end{align}
In this expression $\Delta\theta^\mu$ represents a small increment in either
$F$ or $k$, keeping the other fixed (so, for example, $\theta + \Delta\theta^F$
is interpreted as the probability density at $F+\Delta F, k$).
Following~\cite{Wang2011} we set the expression under the integration to zero
whenever any of the densities was zero for a given $x$. An extended discussion
about the use of non-parametric density estimation (such as Kernel Density
Estimations) and using finite difference schemes in the computation of the
Fisher information is given in~\cite{HarShemesh2015}.

Since our premise is that the distribution of blob sizes is typical to
the different patterns, we use the Shannon entropy, defined as:
\begin{equation} 
  \label{eq:shannon_entropy} 
  H[p](\theta) \equiv \int p(x;\theta) \ln \frac{1}{p(x;\theta)} dx.  
\end{equation}
as a way to validate our method. The Shannon entropy~\eqref{eq:shannon_entropy}
is a measure of uncertainty in the outcome of a random variable which is
distributed according to $p(x;\theta)$. In our case it represents our
uncertainty about the size of one of the blobs in the pattern. In the stable
spots pattern, for example, the uncertainty is relatively low and so is the
Shannon entropy. At the stripe pattern it is high, since any given blob can
belong to a short or a long stripe. This gives us a sort of an order parameter
which, however, is not necessarily related to a change in the symmetry of the
system. This allows us to compare the predictions from the Fisher information
to that of the Shannon entropy. We will typically expect a line of high Fisher
information where the Shannon entropy undergoes a large change.

%%%%%%%%%%%%%%%%%%%%%%%%%%%%%%%%%%%%%%%%%%%

\section{Results}
\label{sec:results}

\subsection{Two dimensional phase map}
\label{sub:two_dimensional_phase_map}

\begin{figure}
  \includegraphics[width=0.45\textwidth]{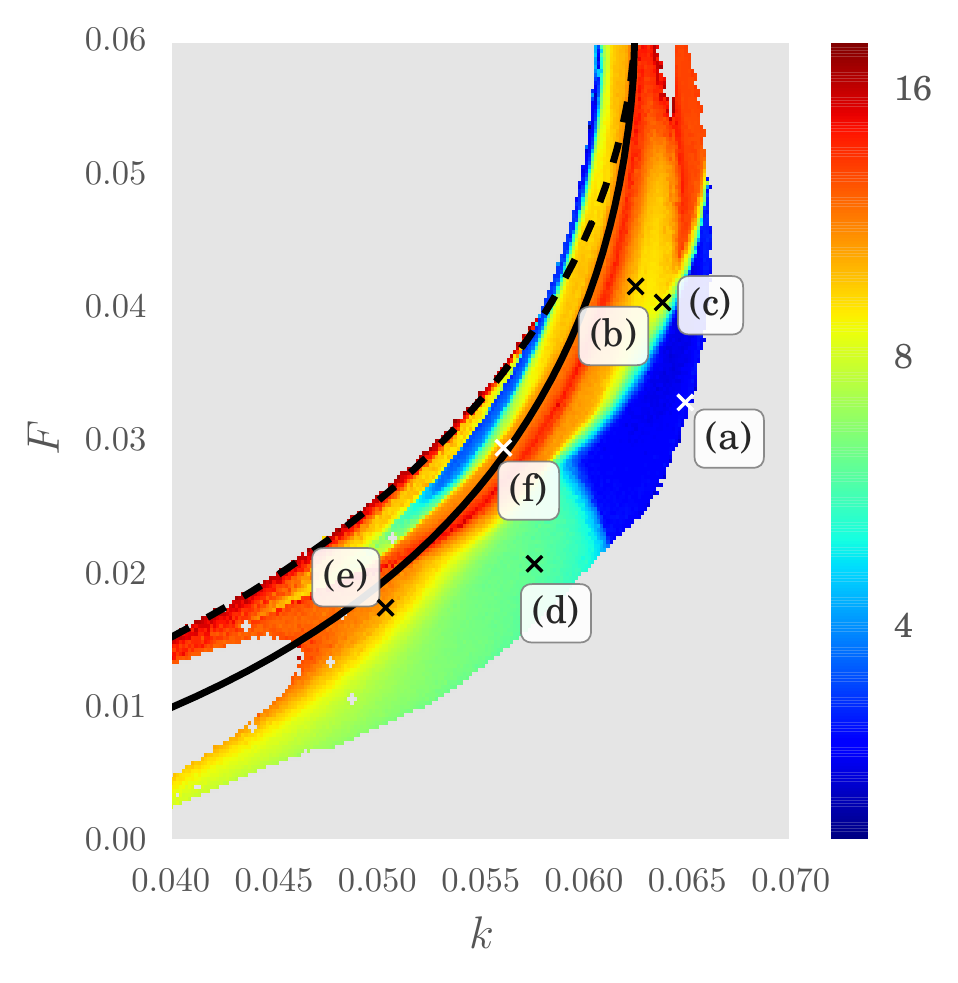}
  \caption{Shannon entropy of the blob size distribution.
    The black continuous line is the saddle-node bifurcation curve and the
    dashed line is the Hopf bifurcation. The Shannon Entropy was calculated for a
    simulation on an $800 \times 800$ grid. The location of the different
    patterns from Fig.~\ref{fig:patterns} are indicated with $\times$. The
    color-bar is in logarithmic scale. Gray indicates areas with no
  inhomogeneous patterns.}
  \label{fig:entropy_patterns}
\end{figure}

To get an overview of where in parameter space each patterns is located, we
first plot the Shannon entropy of the blob-size PDF. The Shannon Entropy map
for a simulation grid of $800 \times 800$ is shown in
Fig.~\ref{fig:entropy_patterns}. The Shannon entropy shows a clear demarcation
between different regions in the parameter space near the saddle-node
bifurcation curve. There are two areas with low entropy (blue regions, one
above and one below the saddle-node curve). Both blue areas have stable
localized spots of roughly the same size, but only the spots in the area
containing the (a) pattern are created through self replication. The other
spots are created with a different mechanism, see the Supplemental Material at
for movies showing the evolution of these two patterns. When we move from the
area containing (a) to the area containing (d) there is a well defined boundary
where the blue region ends and where the spatio-temporal chaotic regime begins.
The Shannon entropy increases slowly as one moves towards lower values of $F$
and $k$ since the size of the spots becomes less certain and since half formed
spots are counted as blobs with smaller and more variable areas. We refer to
this transition as ``Transition I''. A second interesting transition occurs
when moving from (a) to (b). In this transition the spots from pattern (a) mix
with the stripes of pattern (b) to a varying degree, until there are no more
spots and only stripes remain. As we can see the Shannon entropy increase is
much steeper in this transition, since the blobs turn into stripes of wildly
different lengths.  We will refer to this transition as ``Transition II''.
Examples of transitions I and II can be seen in Figs.~\ref{fig:trans1}
and~\ref{fig:trans2} respectively.

\begin{figure*}[ht!]
  \subfloat[$G_{FF}$ component.\label{fig:G_FF}]{%
    \includegraphics[width=0.3\textwidth]{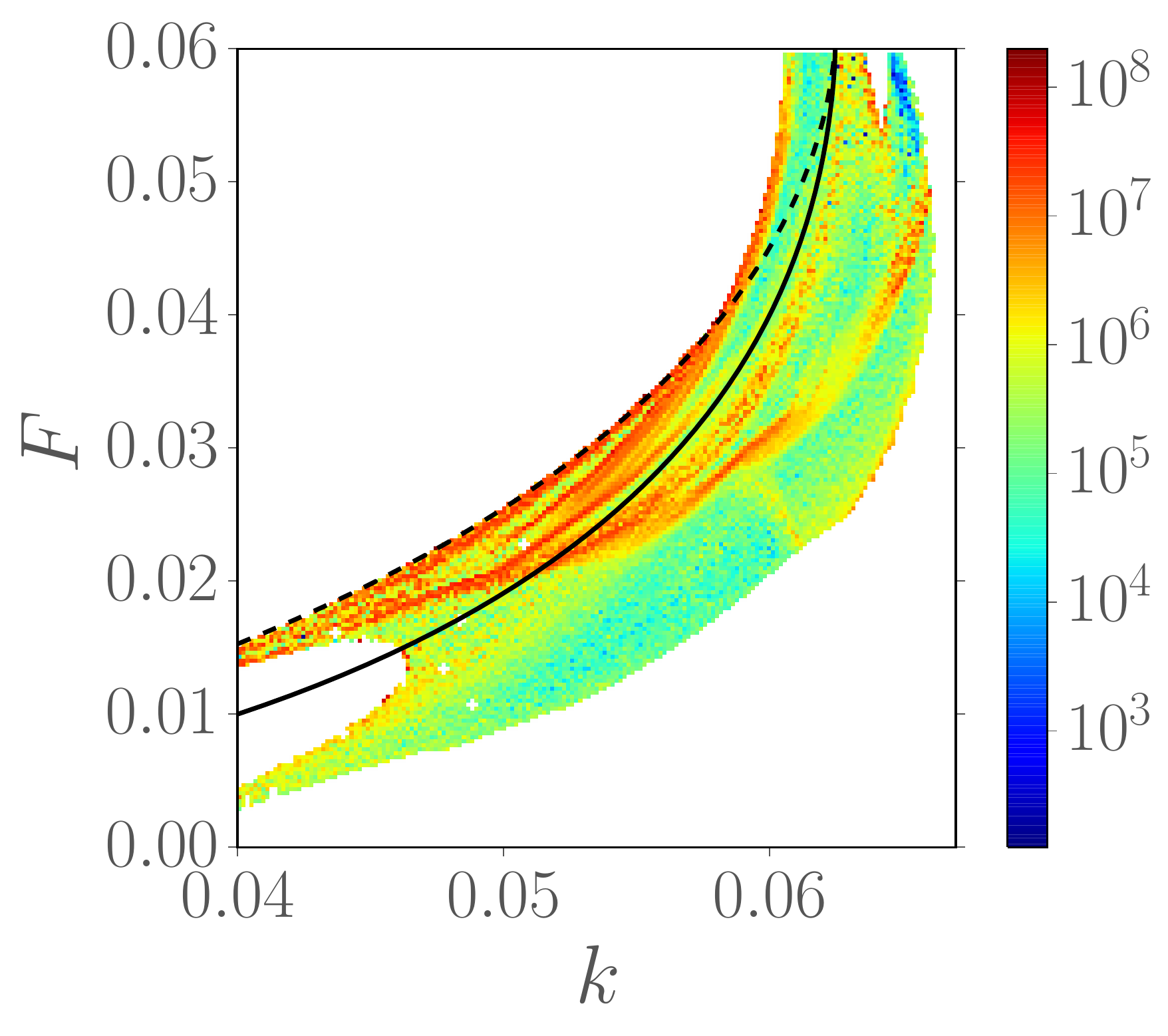}
  }
  \subfloat[$G_{kk}$ component.\label{fig:G_kk}]{%
    \includegraphics[width=0.3\textwidth]{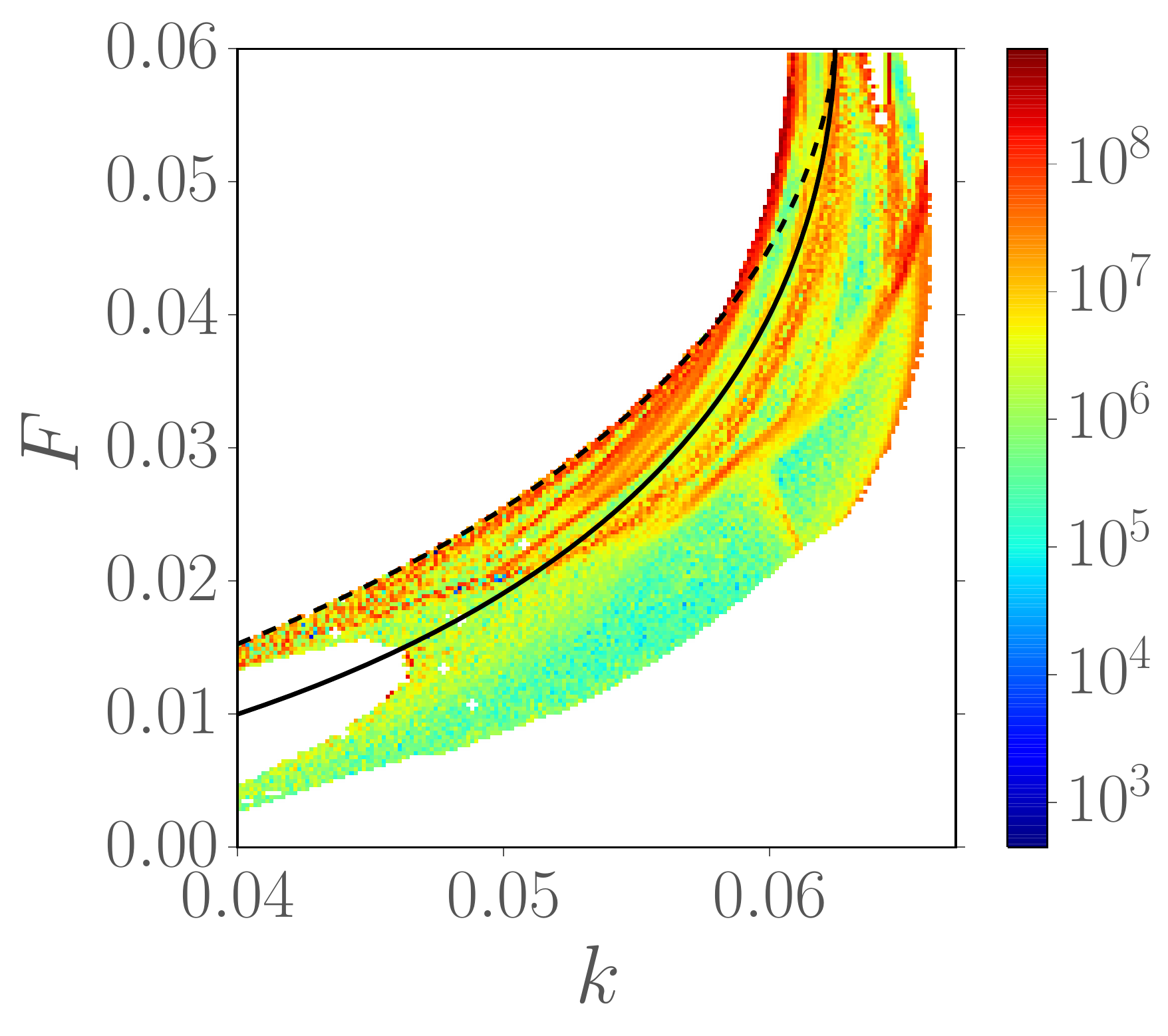}
  }
  \subfloat[$G_{Fk}$ component.\label{fig:G_Fk}]{%
    \includegraphics[width=0.3\textwidth]{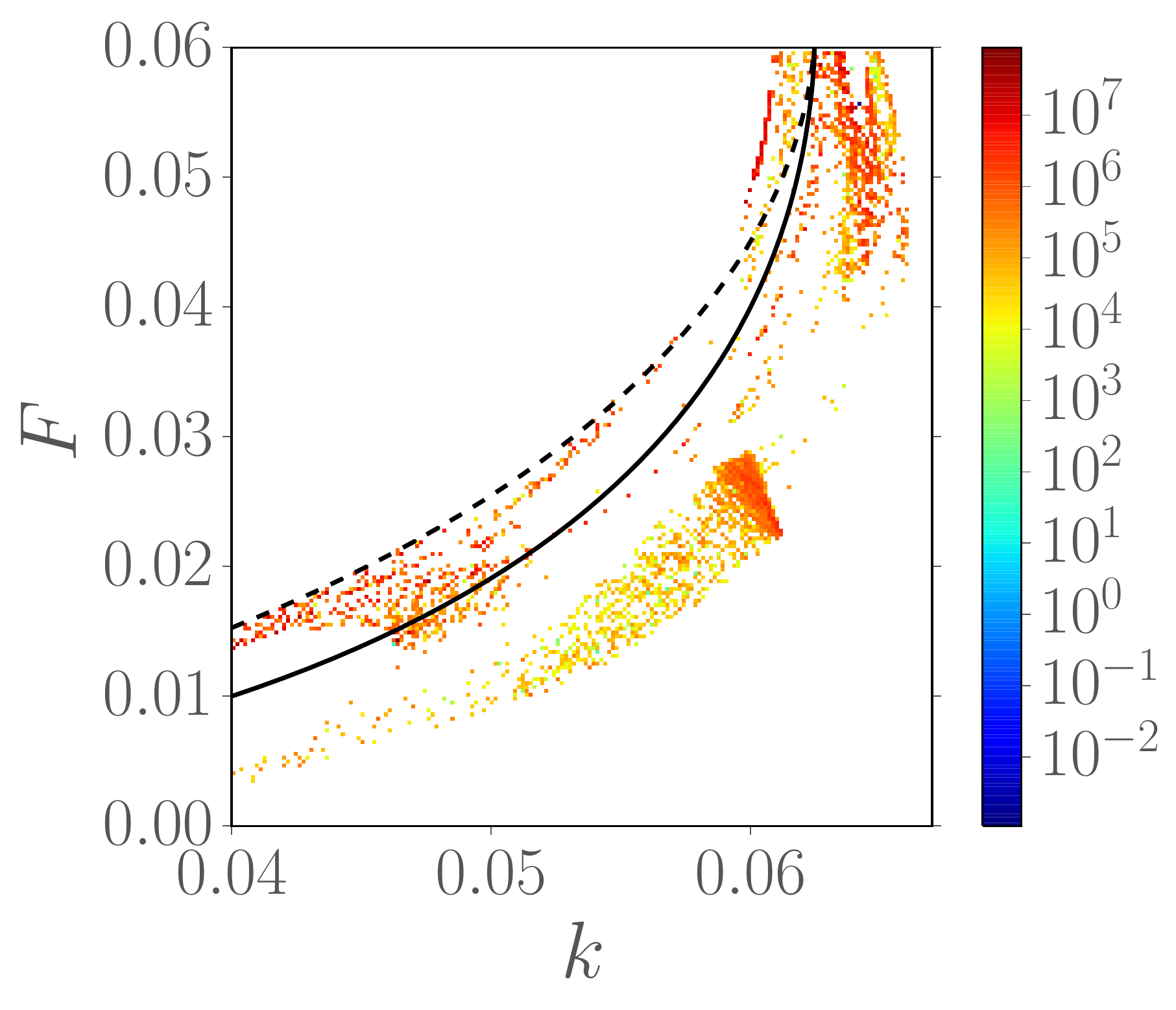}
  }\\
  \subfloat[$\mathrm{Tr\ }G$.\label{fig:trace}]{%
    \includegraphics[width=0.3\textwidth]{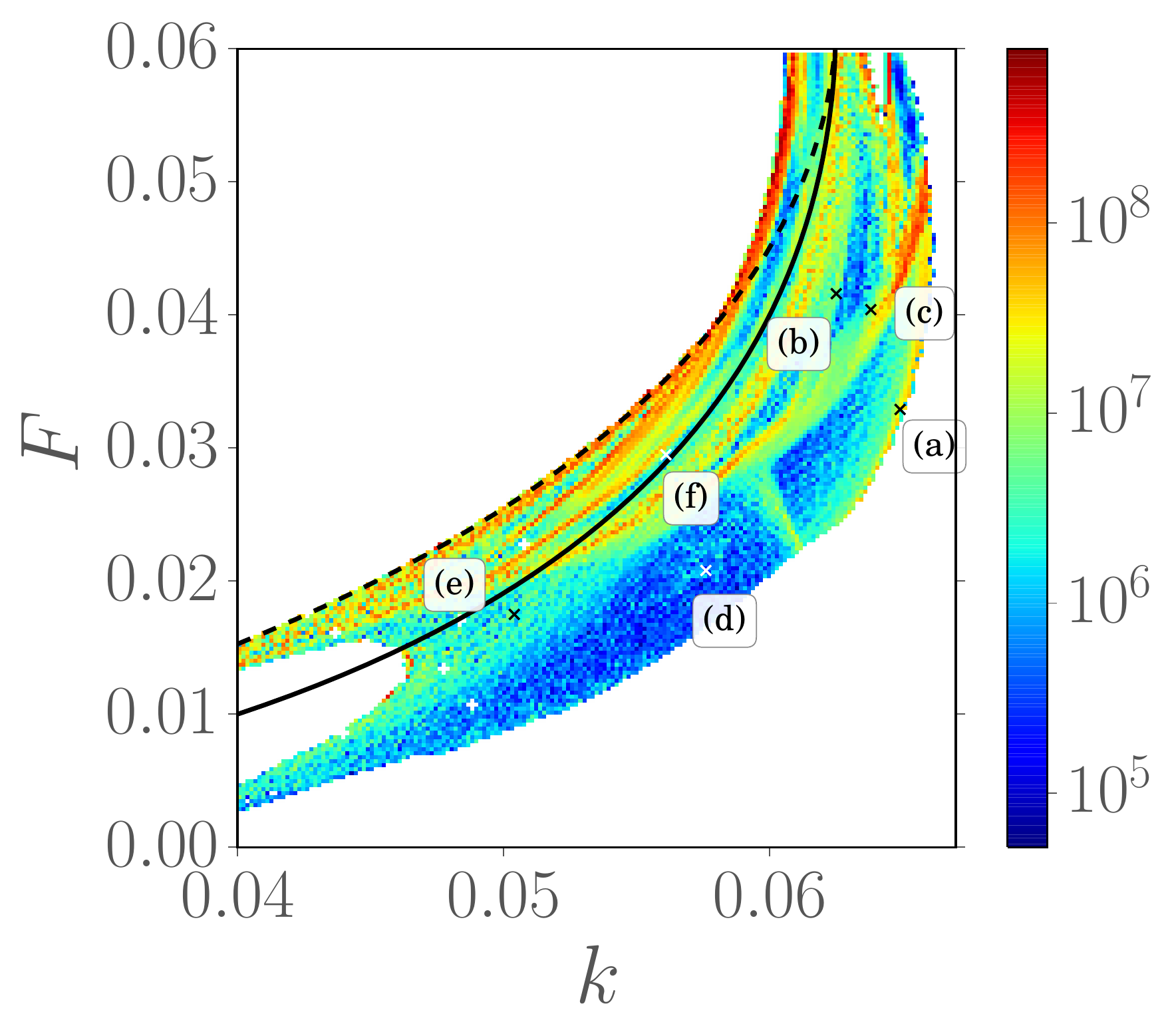}
  }
  \subfloat[Largest eigenvalue.\label{fig:G_ev1}]{%
    \includegraphics[width=0.3\textwidth]{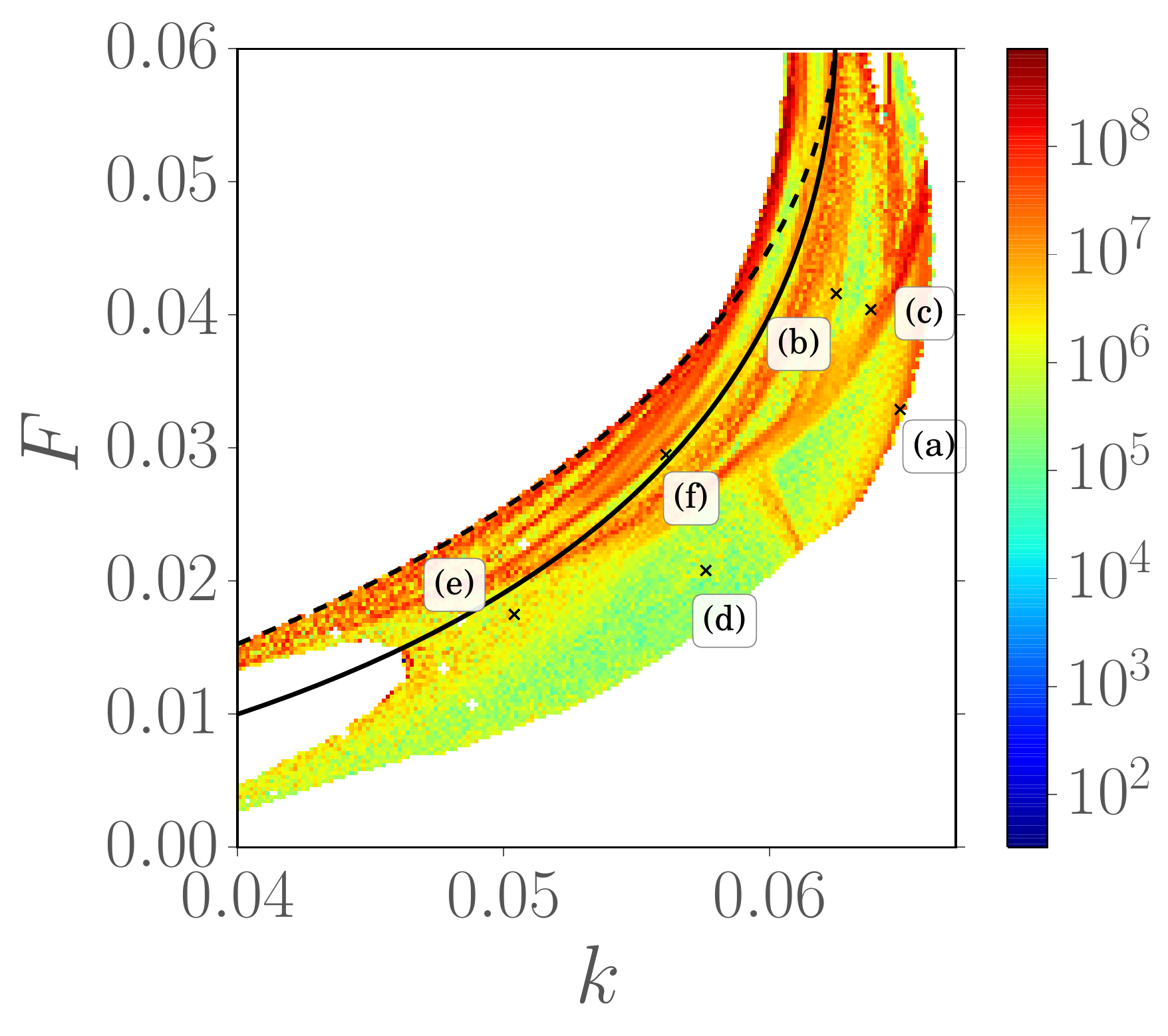}
  }
  \subfloat[Second eigenvalue of $G$.\label{fig:G_ev2}]{%
  \includegraphics[width=0.3\textwidth]{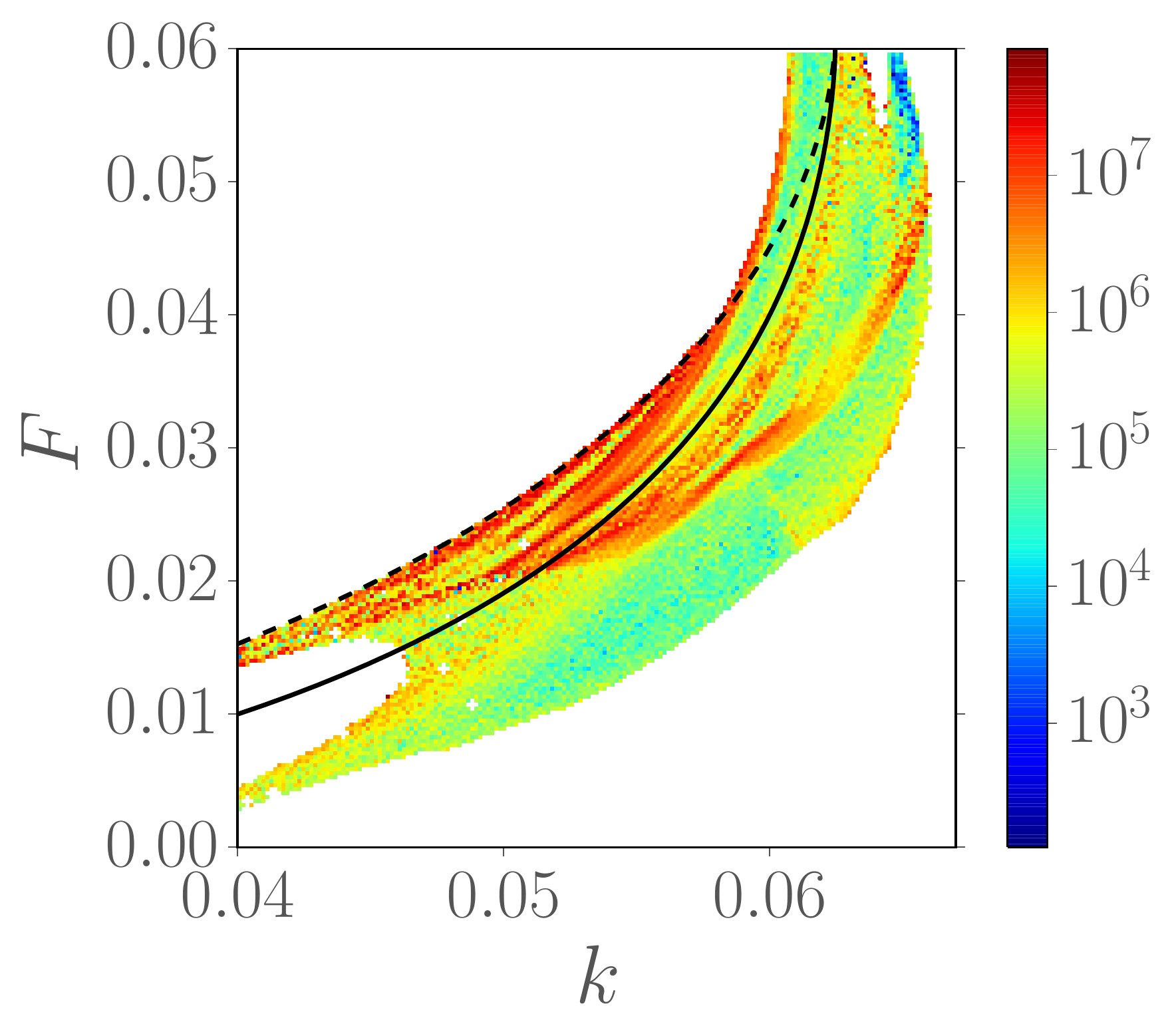}}
  \caption{The different components of the Fisher information matrix obtained
    for simulations with grid size $800\times 800$ and periodic boundary
    conditions. As can be seen from~\ref{fig:G_Fk}, the off diagonal component is
    mostly negligible. The color in all figures is in logarithmic
  scale.}\label{fig:fisher}
\end{figure*}

The different components of the Fisher information matrix and its trace and
eigenvalues are plotted in Figure~\ref{fig:fisher}. At each point we computed
the eigenvalues and plotted separately the larger eigenvalue
(Fig.~\ref{fig:G_ev1}) and the smaller
one (Fig.~\ref{fig:G_ev2}).

The main features of the Fisher information maps we see in
Fig.~\ref{fig:fisher} are that there are many curves with high values of the
Fisher information (``ridges'') which separate areas of lower FI\@.
Most ridges follow the general curve of the saddle-node bifurcation line,
except for the noticeable ridge separating the self-replicating spots pattern
region (a) and the spatio-temporal chaotic area (d) (i.e.\ the ridge
representing Transition I).  Almost all versions of the Fisher map show this
ridge, but with a varying degree of clarity. Transition II is also clearly 
represented by a ridge which appears in almost all representations of the
FIM (except for the $G_{Fk}$ component). This is the rather wide ridge where
pattern (c) resides.

\subsection{One dimensional transitions}
\label{sub:one_dimensional_transitions}

We constructed the two dimensional phase map by simulating the different
patterns that appear in all the different parameter values we were interested
in.  Often we do not have complete knowledge of the surrounding patterns and
only observe one instance of the system at a time. In these cases, it can be
useful to look at a one dimensional plot of the Fisher information along a
given trajectory. This can happen, for example, when we observe a natural
system over time. Then we could treat the observation time as the parameter
$\theta$ and compute how much we know about the observation time from the
distribution $p$.  As an example we plot the Shannon Entropy and the Fisher
information interval along two trajectories. One crossing Transition I and the
other Transition II\@.  We define the Fisher information interval in analogy with
special relativity~\cite{DInverno1992}:
\begin{equation}
  \label{eq:interval}
  ds^2 = G_{\mu\nu} d\theta^\mu d\theta^\nu.
\end{equation}
It is an invariant measure that represents the squared distance along the path
described by $d\theta^\mu$. To compute the interval we first defined the start
and end of the path $\mathbf{p}_1 = (k_1, F_1)$ and $\mathbf{p}_2=(k_2, F_2)$
respectively, then computed the unit vector connecting $\mathbf{p}_1$ and
$\mathbf{p}_2$: $\mathbf{d\hat{r}} =
(\mathbf{p}_2-\mathbf{p}_1)/||\mathbf{p}_2-\mathbf{p}_1||$. And finally the
vector $d\theta^\mu = (dk, dF) \cdot \mathbf{d\hat{r}}$. $dk$ and $dF$ are the
lattice spacings in parameter space and $||\cdot||$ is the usual norm in
$\mathbb{R}^2$. 

\subsubsection{Transition I: self replicating spots to spatiotemporal chaos}

\begin{figure*}[htpb]
  \centering
  \includegraphics[width=0.95\linewidth]{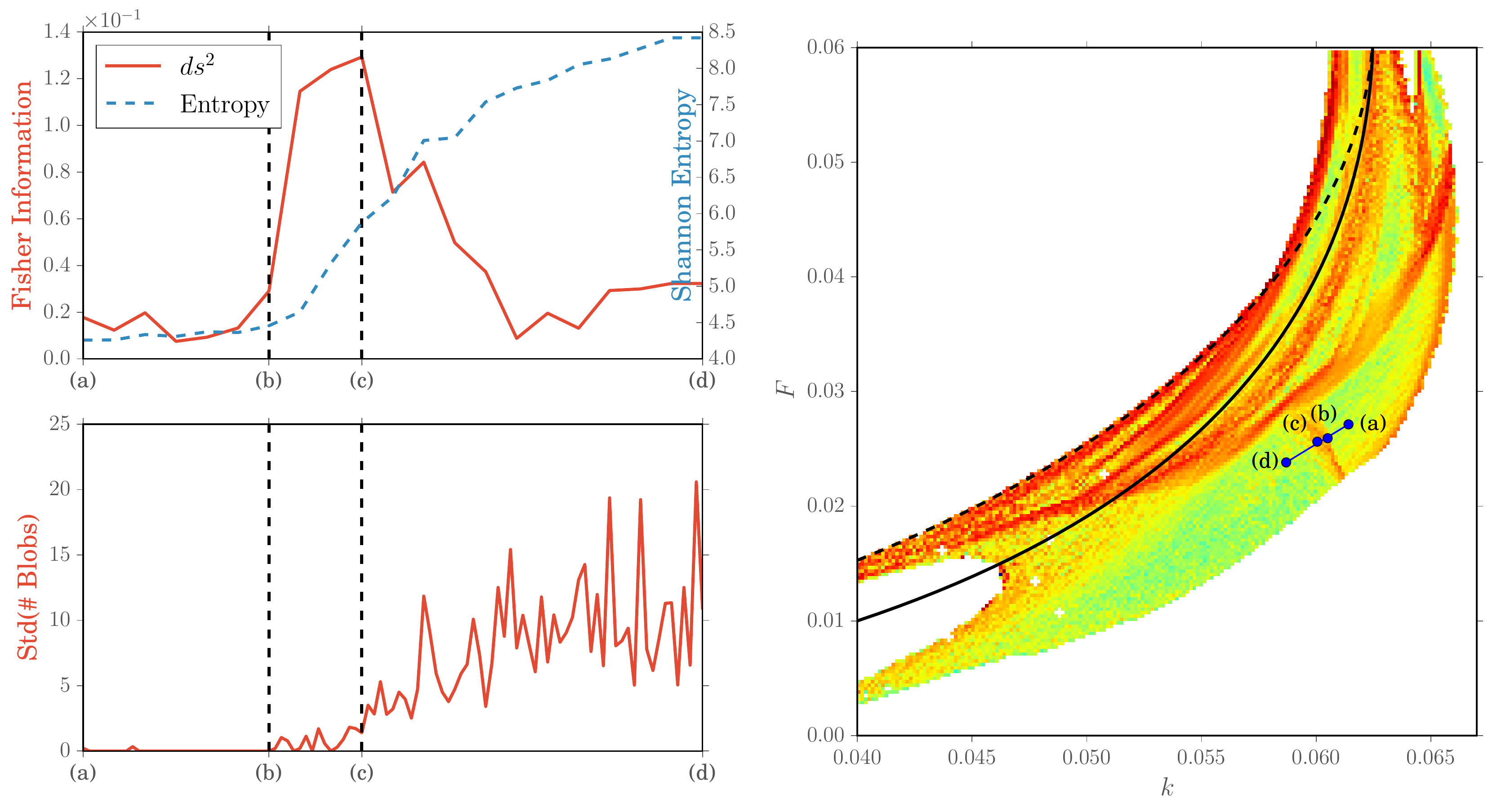}
  \caption{One dimensional snapshot of Transition I, starting from point (a) to
    point (d), crossing the ridge defining the transition. Point (b) represents
    the onset of the rise of the Fisher information and point (c) represents
    the maximum of the Fisher information along the line.  Top left figure - the
    line element $ds^2 = G_{\mu\nu}d\theta^\mu d\theta^\nu$ connecting (a) and (c), and
    the Shannon entropy along the line.  Bottom left plot represents the standard
    deviation in the blob count from a continued simulation along the line, as
    explained in the main text.  The vertical dashed line in both plots
    represents the position of (b) and (c).  The plot on the right is the same
    line element $ds^2$ plotted for the entire simulation region. The blue line
    is the line plotted on the left. Points (a), (b), (c), and (d) are also
  indicated.}\label{fig:trans1}
\end{figure*}

The first one dimensional transition we analyze is Transition I, which we
define between the stable, self-replicating, spots and the chaotic spots. One
such trajectory crossing the transition is plotted in Fig.~\ref{fig:trans1}.
Point $(a)$ (at parameter values $k = 0.06141, F = 0.027136$) is located within
the fixed spots area close to the transition and point $(d)$ (at parameter
values $F = 0.0238191, k=0.05869347$) is located within the chaotic spots area,
also not far from the boundary. On the top left of Fig.~\ref{fig:trans1} we
plot the Fisher interval going from point $(a)$ to point $(d)$. At point $(b)$
the interval starts increasing and at point $(c)$ it reaches a maximum, after
which is slowly decreases again. On the same graph we plot the Shannon entropy
of the blob size PDF\@. The Shannon entropy is lower in the stable phase than
in the spatio-temporal chaos, as can also be seen in
Fig.~\ref{fig:entropy_patterns}.  This is due to the increased uncertainty in
blob sizes in the chaotic regime.  

We inspected the patterns on both sides of the transition visually and as a
function of time and it does indeed seem to capture the transition in the
correct location.  To verify this quantitatively we define an order parameter.
Since the nature of the transition is dynamic (i.e.\ the time dependence of the
pattern is different in both phases) we simulate the system further in time.
We divide the trajectory from $(a)$ to $(d)$ to $100$ points at each point run
the simulation for $50,000$ warm-up time steps for it to reach the state in
which we computed the Fisher information. We then continue the simulation an
additional $2,000$ time steps, computing the blob-count every $20$ time steps.
From these samples we compute the standard deviation of the blob count.  The
rationale is that for the stable spots the variability of blob count should be
zero and at the chaotic regime it is non-zero. This is plotted on the bottom
left in Fig.~\ref{fig:trans1}, below the Fisher interval plot. We also plotted
vertical lines that indicate the position of points $(b)$ and $(c)$.  Because
of the demanding computation time we performed the validation run on a $400
\times 400$ grid as opposed to the $800 \times 800$ grid from which the Fisher
interval is computed. The blob-count indeed remains constant between $(a)$ and
$(b)$.  It then gradually becomes more variable as we go deeper into the
chaotic regime.  The increase in blob-count variability coincides with the rise
in Fisher interval which we interpret as a verification that indeed the
transition begins at this point. 

\subsubsection{Transition II: self-replicating spots to stripes}

\begin{figure*}[htpb]
  \centering
  \includegraphics[width=0.95\linewidth]{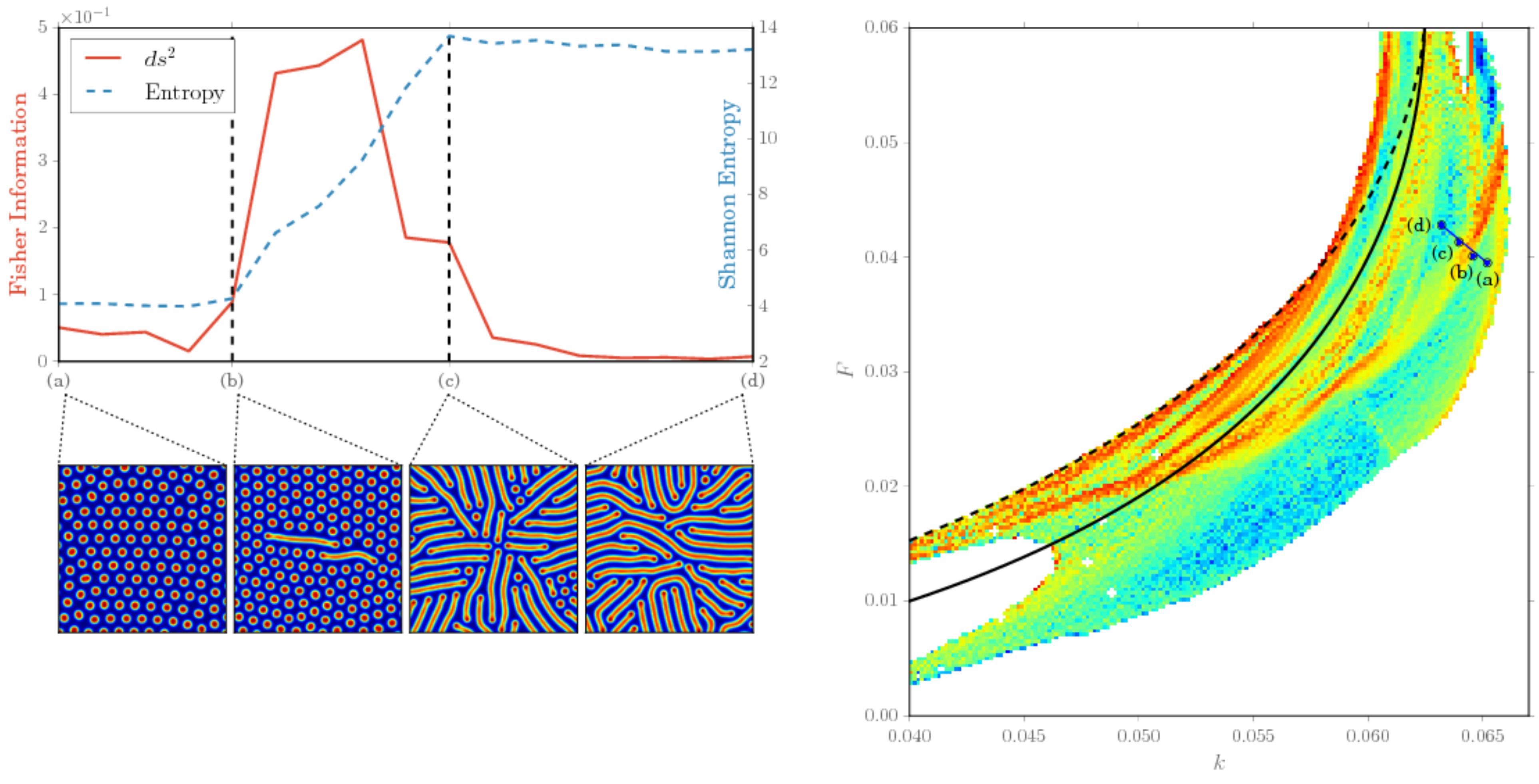}
  \caption{One dimensional snapshot of Transition II\@. Top left is a plot of the
    interval $ds^2=G_{\mu\nu}d\theta^\mu d\theta^\nu$ and the Shannon entropy
    along the line connecting point (a) in the stable self-replicating spots area
    and (d) in the stripes area. Points (b) and (c) represent the onset of increase
    of Fisher information and the point where the Shannon Entropy reached its maximal
    value. The images on the bottom represent zoomed-in depictions of the
  patterns appearing in all indicated points.}
  \label{fig:trans2}
\end{figure*}

The second trajectory we chose crosses what we term Transition II\@. As the
first transition, it starts in the self-replicating spots area (at a different
point $(a)$ which is located at parameter values $k=0.0652$ and $F=0.0395$) and
crosses to the area of the stripe patterns to point $(d)$ which is located at
$k=0.0632, F=0.0428$. Along the trajectory the patterns evolve from completely
spots to a mixed spot-stripe region and finally reach a region without any
spots. Again we plot the trajectory in Fig.~\ref{fig:trans2}, along with the
Fisher interval and Shannon entropy (top left). The Fisher interval clearly
increases between points $(b)$ and $(c)$ which this time designate visually
selected points of the start and end of the region where the Fisher interval is
high. The Shannon entropy again increases across the transition and is higher
in the stripe phase in comparison with the self-replicating spot phase. Below
the Fisher interval and Shannon entropy plots we draw a part of the patterns
that appear at points $(a)$ through $(d)$. This helps to provide a visual
verification of the position of the transition. Point $(b)$ is the first point
with stripes appearing together with the spots and point $(c)$ is one of the
last with spots.

\subsubsection{Shannon Entropy and Fisher Information}

It is tempting to compare the Shannon entropy and Fisher information in terms
of how well they capture the transition. Both can be used as indicators, the
Shannon entropy indicates a transition when its average value changes
significantly and the Fisher information by rise and subsequent decrease in its
value. We would like to note that, following the discussion earlier, the Fisher
information acts as a susceptibility measure (as the magnetic susceptibility
would act in an Ising model) and the Shannon entropy as an order parameter (as
the magnetization would in an Ising model). We can however hypothesize that it
is possible to find a transition between patterns in which the Shannon entropy
is constant but the Fisher information peaks. For example, if the average blob
size changes abruptly but the variance in blob sizes remains fixed. In that
case the Shannon entropy would not catch the transition but the Fisher
information would.

\subsection{Effect of the simulation grid size}
\label{sec:grid_size}

The size of the simulation grid, which we varied in our experiments from $256
\times 256$ up to $1600 \times 1600$ has a large effect on the resulting phase
map. Small grid sizes result in a fairly noisy Fisher map, whereas larger ones
give a smoother map with better defined ``ridges'' which demarcate the
different phases. We suspect this is a result of too few blobs appearing in the
smaller grids. Because simulating large grid sizes requires considerable
computation time (the $1600 \times 1600$ took about a month on a cluster running
on approximately $800$ cores simultaneously), one has to find a balance between
accuracy and computational time. We found that a grid size of $800 \times 800$
provided good results. 

\begin{figure*}[htpb]
  \centering
  \subfloat[Grid $256\times 256$]{%
  \includegraphics[width=0.45\linewidth]{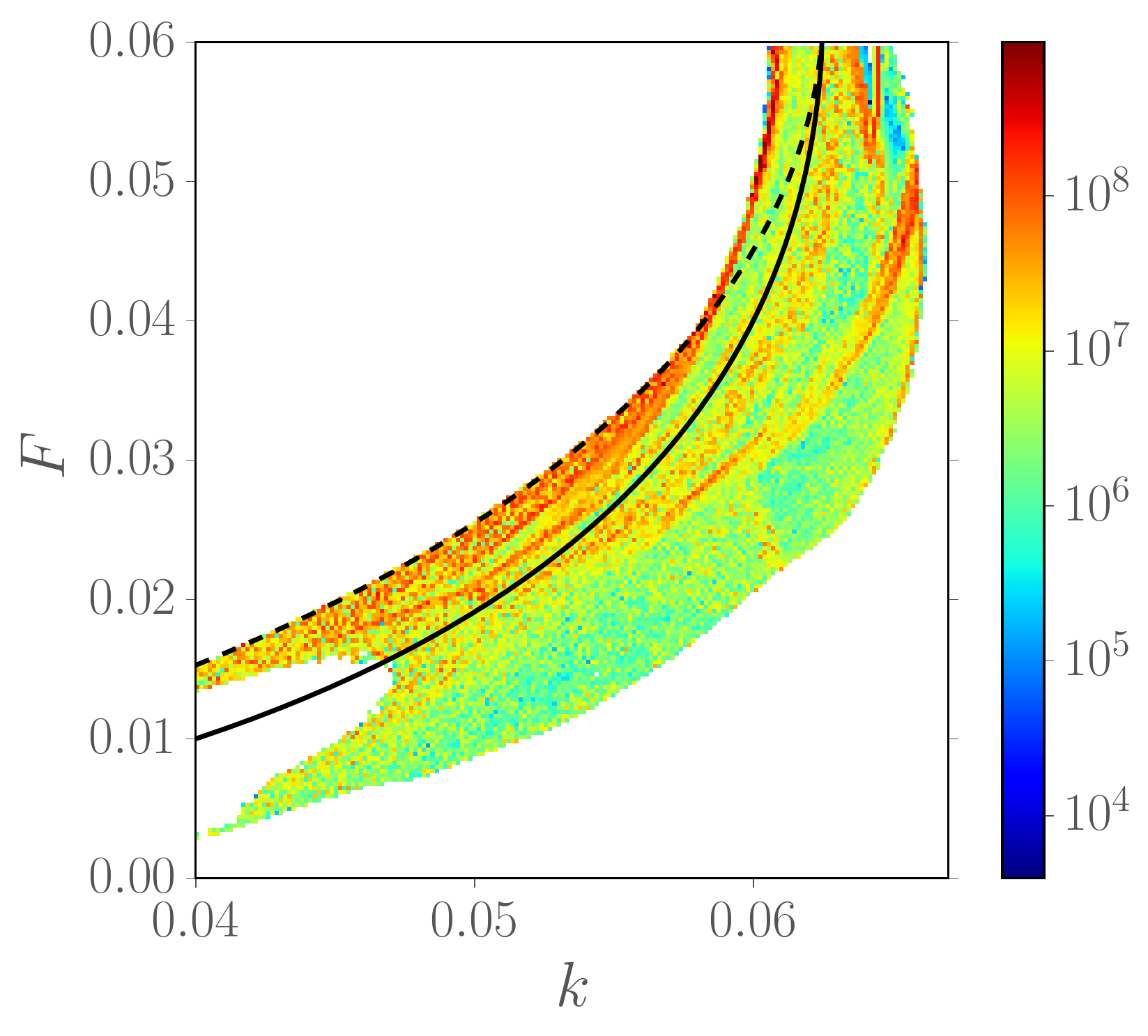}}
  \subfloat[Grid $400\times 400$]{%
  \includegraphics[width=0.45\linewidth]{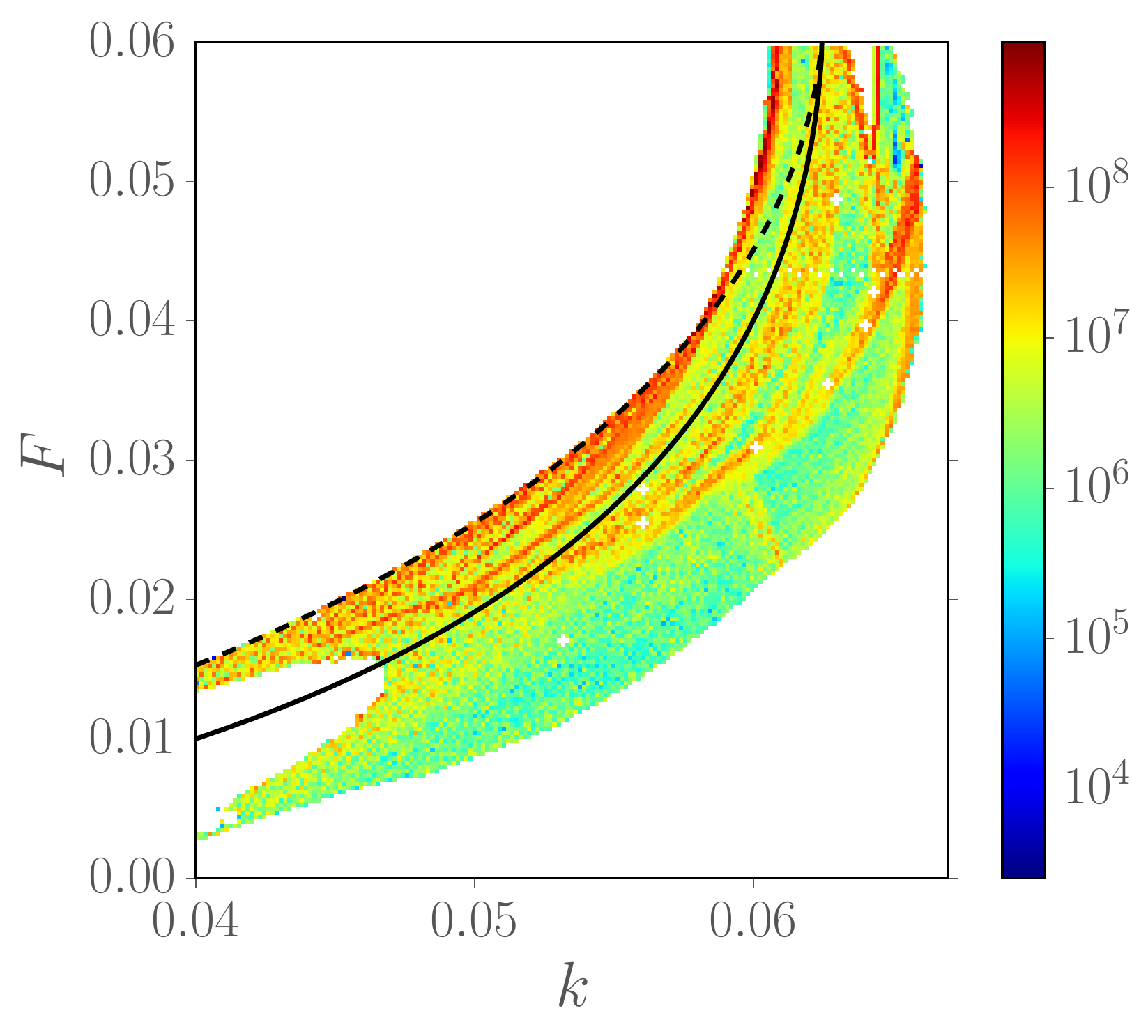}}\\
  \subfloat[Grid $800\times 800$]{%
  \includegraphics[width=0.45\linewidth]{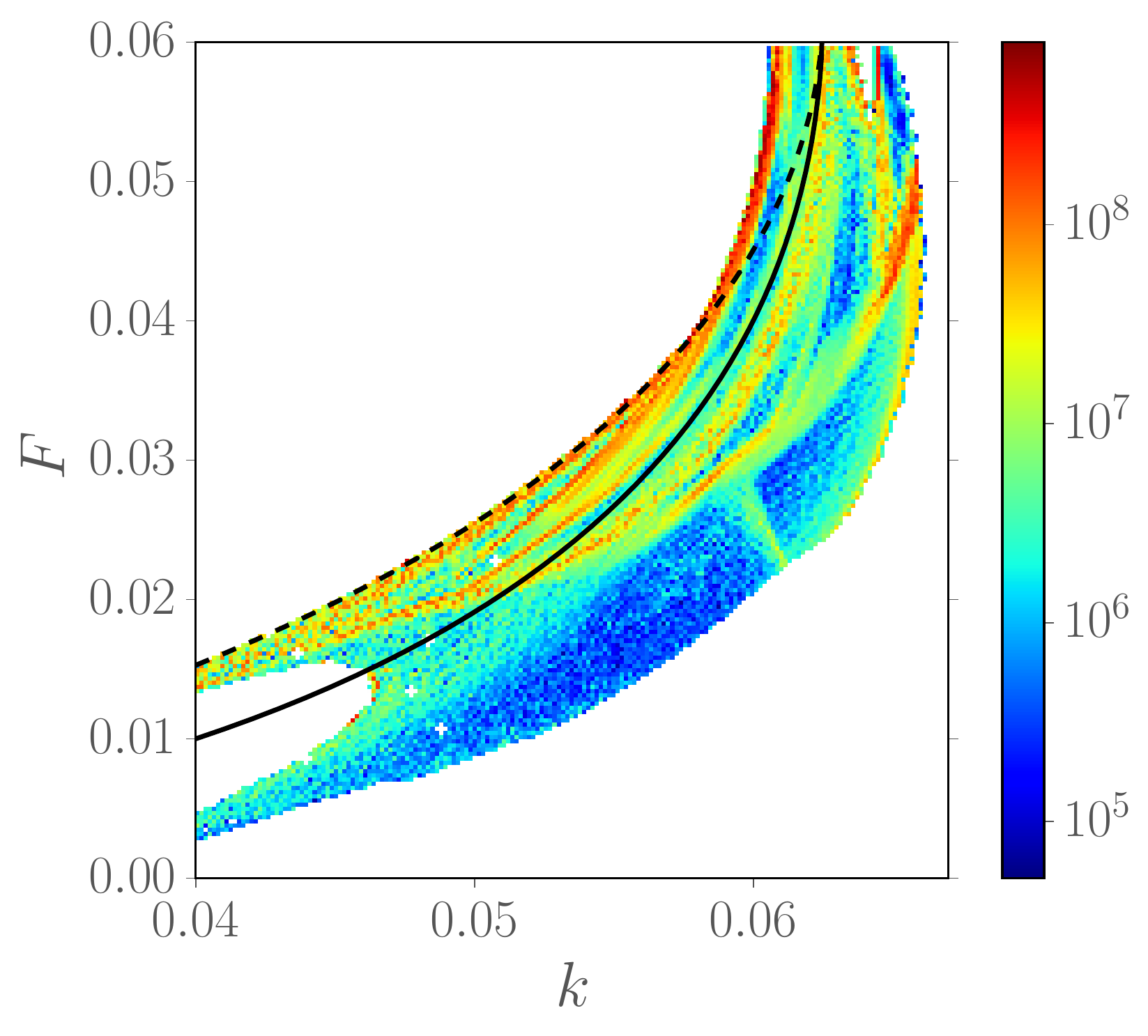}}
  \subfloat[Grid $1600\times 1600$]{%
  \includegraphics[width=0.45\linewidth]{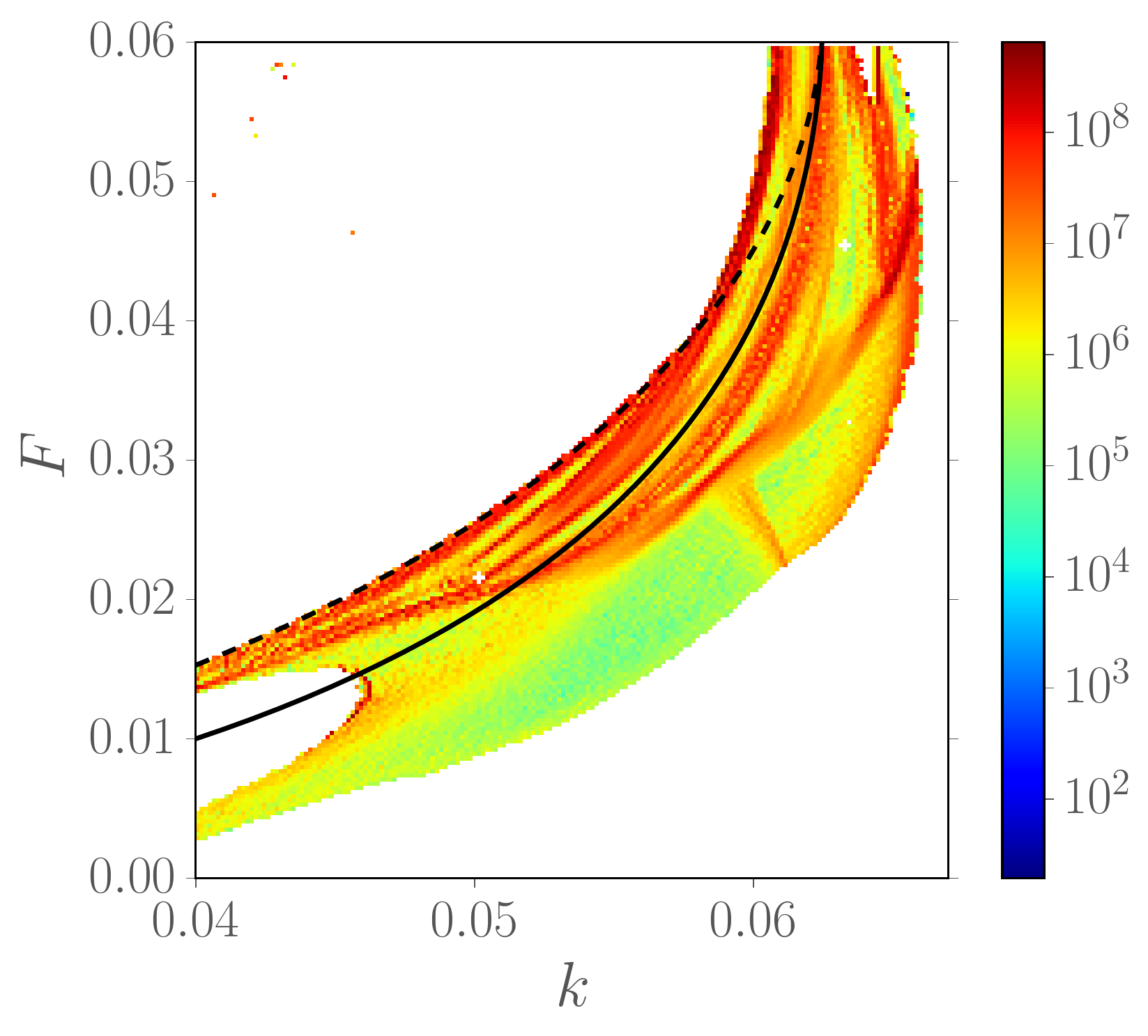}}
  \caption{Comparison of different grid sizes. Trace of the FIM.}
  \label{fig:grid_sizes}
\end{figure*}

In order to compare the results obtained for each of the grid sizes, we plot
the trace of the Fisher information matrix obtained from the computation of the
different grid sizes. This is presented in Fig.~\ref{fig:grid_sizes}.  At $256
\times 256$ there is quite some noise from large peaks of the Fisher
information. The $400 \times 400$ grid already provides a much better
resolution of the ridges, which improves even more at the $800 \times 800$
grid. The $800\times 800$ grid also shows the highest contrast in Fisher
information between ``ridge'' and ``valley'' regions.  As we go to the largest
grid size we used, $1600 \times 1600$, it seems that some of the well-defined
ridges become less well defined. We suspect that this might happen because the
system did not have sufficient time to reach equilibrium, even though each
simulation was run for $200,000$ time steps. 

%%%%%%%%%%%%%%%%%%%%%%%%%%%%%%%%%%%%%%%%%%

\section{Methods}
\label{sec:methods}

The Fisher information is obtained by performing simulations on the parameter
range where inhomogeneous patterns appear in the Gray-Scott model. The
parameter space was divided into an evenly spaced $200\times 200$ grid with $F
\in [0, 0.06];\; k \in [0.04, 0.07]$. The diffusion coefficients were held
constant at $D_u = 0.16;\; D_v = 0.08$ such that $D_u/D_v=2$. We performed a
simulation at each point of parameter space, starting with identical initial
conditions (same seed) and repeated for the same number of time steps
(depending on the simulation grid size). The simulation was started with an
initial condition of the red state $(u,v) = (1,0)$ with a finite perturbation
in the form of a $20 \times 20$ square in the center of the simulation grid in
the state $(u, v) = (0.5, 0.25)$ and an additional Gaussian noise with an
amplitude of $0.05$ covering the entire simulation grid. This initial state was
then evolved by integrating numerically Eq.~\eqref{eq:gs_diff} using an Euler
scheme until the final state was reached. We repeated the experiment with
different simulation grid sizes, ranging between $200 \times 200$ up to $1600
\times 1600$.  The simulation times ranged from $50,000$ time steps (for the
smallest grid sizes) to $200,000$ for the $1600 \times 1600$ grid. This was
chosen such that the self-replicating stable spots will fill in the entire
simulation window.  The simulations were performed on the Lisa cluster run by
SurfSara~\footnote{\url{https://userinfo.surfsara.nl/systems/lisa/description}}.
The python code to perform the simulation is based on the code found
at~\footnote{\url{http://www.loria.fr/~rougier/teaching/numpy/scripts/gray-scott.py}}.

For each simulation we extracted the PDF by following the steps described in
Section~\ref{sub:probabilistic_description}. We used the Python package
\verb;SimpleCV; for the binarization and blob detection of the images and
\verb;scipy.stats.gaussian_kde; for the computation of the PDF from the blob
sizes. The computation of the Fisher information from the PDF followed the
description in~\cite{HarShemesh2015} and the code for this computation is
available online
at~\footnote{\url{http://uva.computationalscience.nl/research/software/npfi}}.  As
mentioned in Sec.~\ref{sec:results} we also computed the Shannon entropy for
each PDF we obtained. This was done by simple integration of the PDF using
Eq.~\eqref{eq:shannon_entropy} and the Python function
\verb;scipy.integrate.quad;.

In addition to the Gaussian KDE, we used the novel density estimation method
DEFT~\cite{Kinney2014a}, and tried two different ways to integrate
Eq.~\eqref{eq:fisher_finite_diff} - once as it is written in
Eq.~\eqref{eq:fisher_finite_diff} and once by first performing the
differentiation of the logarithm (replacing $\partial_\mu \ln p$ with
$[1/p]\partial_\mu p$) before defining the finite difference scheme. As is
described in~\cite{HarShemesh2015}, the first method yields better results
for Gaussian KDE and the second for DEFT\@. We eventually used the results
obtained from the Gaussian KDE rather than the DEFT ones because finding the
correct parameters for DEFT for all parameter values was difficult, since both
the range of blob sizes and the number of blobs in each simulation in each
parameter value varied too much. 

\section{Conclusions}
\label{sec:conclusions}

In this paper we explore the use of the Fisher information matrix to capture
transitions between different patterns in the Gray-Scott model. The use of
Fisher information for this purpose is inspired by an analogy with phase
transitions in thermodynamic systems and follows the work by Prokopenko
\emph{et al.}~\cite{Prokopenko2011}. The main purpose of the study was to test
whether such a description is feasible in a system whose probabilistic
description is not derived from the microscopic dynamics and where no
statistical model is assumed. We find that at least in the case of the GS model
with the choice of blob-size PDF as a probabilistic description, this seems to
be indeed possible.  

The main difficulty in using this approach to produce the entire phase map is
that it is computationally very demanding. The very smooth phase map only
really appears at grid sizes of $800 \times 800$ which requires long
computation runs. A more conceptual difficulty is that once we obtain the maps,
their exact interpretation is not trivial. The value of the Fisher information
ranges over many orders of magnitude (between $10^3$ and $10^9$) and there is
no theoretical value to compare with. 

The strength of the method lies in that it manages to capture various types of
transitions with the same metric. We did not have to define specific order
parameters for the different patterns. It also captures our intuitive
understanding of a ``phase transition'' as a large change in the statistical
properties of a system when the underlying ``external'' parameters are changed.

In the one-dimensional cases, it seems that the Fisher information does indeed
capture essential features of ``pattern transitions'' in that it gives a clear
signal in the form of a peak between two regions of low Fisher information. It
is especially interesting to note that the Fisher information predicts an exact
position for the transition to spatio-temporal chaos, as depicted in
Fig.~\ref{fig:trans1}, since this transition is dynamic in nature.  Further
mathematical analysis similar to the one done
in~\cite{Nishiura1999,Nishiura2001} for the two-dimensional Gray-Scott model
might be able to confirm the exact position of the transition line and compare
it to the one detected by the FI. 

Our results suggest that Fisher information can indeed serve as a generalized
susceptibility measure in the study of complex systems. Because very few
assumptions about the underlying dynamics are necessary it is an ideal tool
for detecting different regimes even in the absence of an assumed statistical
model.

\begin{acknowledgments}

OHS would like to thank Lisa Jenny Krieg from the University of Amsterdam for
many enlightening discussions and the critical reading of the manuscript. The
research leading to these results has received funding from the European Union
Seventh Framework Programme (FP7/2007-2013) under grant agreement numbers
317534 and 318121. RQ and PMAS acknowledges the financial support of the Future
and Emerging Technologies (FET) programme within the Seventh Framework
Programme (FP7) for Research of the European Commission, under the
FET-Proactive grant agreement TOPDRIM, number FP7-ICT-318121. AgH wishes to
acknowledge partial funding by the Russian Scientific Foundation, under grant
\#14-11-00826. PMAS wishes to acknowledge partial funding by the Russian
Scientific Foundation, under grant \#14-21-00137. 

\end{acknowledgments}

\end{document}